\documentclass{IEEEtran}

\usepackage{cite}
\usepackage{braket}
\usepackage{amsmath,amssymb,amsfonts}
\usepackage{algorithmic}
\usepackage{graphicx}
\usepackage{textcomp}
\def\BibTeX{{\rm B\kern-.05em{\sc i\kern-.025em b}\kern-.08em
    T\kern-.1667em\lower.7ex\hbox{E}\kern-.125emX}}

\usepackage{amsfonts}
\usepackage{amsmath}
\usepackage{amssymb}
\usepackage{array}
\usepackage{mathrsfs}
\usepackage{leftidx}
\usepackage{graphicx}
\usepackage{epstopdf}
\usepackage{dcolumn}
\usepackage{bm}
\usepackage[usenames]{color}
\usepackage{colortbl,booktabs}
\usepackage{subfigure}
\usepackage{cite}
\usepackage{multirow}
\usepackage{amsthm}

\usepackage{ulem}

\usepackage{cases}
\usepackage{soul}
\usepackage[switch]{lineno} 

\begin{document}

\title{Quantum coherent feedback control with photons}

\author{Haijin Ding and  Guofeng Zhang
\thanks{This research is partially supported by Hong Kong Research Grant Council (Grants Nos. 15203619 and 15208418), Shenzhen Fundamental Research Fund, China, under Grant No. JCYJ20190813165207290, National Natural Science Foundation of China under Grant No. 62173288, and the CAS AMSS-polyU Joint Laboratory of Applied Mathematics. (Corresponding author: Guofeng Zhang.)}
\thanks{Haijin Ding was with the Hong Kong Polytechnic University Shenzhen Research Institute, Shenzhen, 518057, China. He is now with the Laboratoire des Signaux et Syst\`{e}mes (L2S), CNRS-CentraleSup\'{e}lec-Universit\'{e} Paris-Sud, Universit\'{e} Paris-Saclay, 3, Rue Joliot Curie, 91190, Gif-sur-Yvette, France (e-mail: dhj17@tsinghua.org.cn). }
\thanks{Guofeng Zhang is with the Department of Applied Mathematics, The Hong Kong Polytechnic University, Hung Hom, Kowloon,  SAR, China, and The Hong Kong Polytechnic University Shenzhen Research Institute, Shenzhen, 518057, China (e-mail: guofeng.zhang@polyu.edu.hk).}
}

\maketitle

\begin{abstract}
The purpose of this paper is to study two-photon dynamics induced by the coherent feedback control of a cavity quantum electrodynamics  (cavity-QED) system coupled to  a waveguide. In this set-up, the two-level system in the cavity can work as a photon source, and the photon emitted into the waveguide can re-interact with the cavity-QED system many times after the transmission and reflection in the waveguide, during which the feedback can tune the number of the photons in and out of the cavity. We analyze the dynamics of two-photon processes in this coherent feedback network in two scenarios: the continuous mode coupling scheme and the discrete periodic  mode  coupling scheme between the waveguide and cavity. The difference of these coupling schemes is due to their relative scales and the number of semi-transparent mirrors for coupling. Specifically, in the continuous mode coupling scheme, the generation of two-photon states is influenced by the length of the feedback loop by the waveguide and the coupling strength between the waveguide and the cavity-QED  system. By tuning the length of the waveguide and the coupling strength, we are able to generate two-photon states efficiently. In the discrete periodic  mode  coupling scheme, the Rabi oscillation in the cavity can be stabilized and there are no notable two-photon states in the waveguide.
\end{abstract}


\section{Introduction}
In recent years quantum feedback control has attracted much attention due to its wide applications in quantum information processing (QIP) such as quantum error correction~\cite{FeedbackError,ErrorFeedback2,cardona2019Error,pan2016stabilizing}, quantum optical fields amplification~\cite{feedbackapplication,song2021amplification}, stabilization of  Rabi oscillation~\cite{2012Stabilizing}, robust stabilization of quantum states~\cite{cardona2020exponential,liang2019exponential,liangRobust}, entanglement generation~\cite{FeedbackEntanglement,ZJL15}, and so on. 
Depending on whether or not the quantum state is measured to yield classical information for  feedback control design, quantum feedback control can be divided into two categories: measurement feedback control and coherent feedback control~\cite{Zhangjing2017,dong2022quantum}. Measurement feedback control has been widely used in the generation of quantum states and error corrections in quantum computation. For example, the three-qubit ``bit-flip" method is the simplest method for correcting error bits~\cite{FeedbackError,ErrorFeedback2}. In measurement feedback control, measuring the quantum state induces external disturbances and normally alters quantum dynamics~\cite{Loyld,QNDscience,RMPMeasurement,Coherentbeat,FeedbackKerr,MeaFBRL,ContinuousMeasFB,OpenMeasFB2022}. On the other hand, in a coherent feedback control network~\cite{gough2009series,JG10,ZJ11,ZJ12,NY14,NY17,CKS17,XPD17,KJN18,GP18,Mabuchi19,PJU+20,AMD20,feedbackapplication,BNC+21,KPS+21,PVB+21,HCS21,Cobarrubia21,WHM+22,dong2022quantum,DRM+22,AtomCoolingExper,IonTrapTwpBitCoheFBPRXQ,twoCavityFBCool,ZD22}, no measurement is involved, and thus the quantum coherence can be preserved during the time evolution of the quantum network.

In a coherent feedback network, the quantized field (e.g., a photon) is scattered from the quantum system (e.g., an atom or atomic ensemble), then it can be redirected to the quantum system to realize the desired evolution~\cite{Mabuchi19,Mabuchi2008,2008H,AtomCoolingExper}. For example, as illustrated in Ref.~\cite{FeedbackRabi}, an atom emits a photon into a semi-infinite waveguide,  the emitted photon is later reflected by the end of the waveguide and interacts with the atom again. Coherent feedback has been realized on various experimental platforms such as superconducting circuits, optical systems, nitrogen-vacancy(NV) centre in diamond, and so on.  For example, in a superconducting circuit, coherent feedback can be used to stabilize the dynamics of the quantum state~\cite{2012Stabilizing} and implement a bistable state~\cite{Bistable}. In an optical system, coherent feedback can enhance the squeezing of the optical beam~\cite{2011ExperimentalSqueezing,YPA20}. For quantum computation based on trapped ions, the real-time coherent feedback control on the target ion can be designed by measuring another correlated ion, and the control efficiency and coherence of the measurement-free target qubit can be enhanced~\cite{IonTrapTwpBitCoheFBPRXQ}. Coherent feedback has also been realized in NV centers~\cite{PaolaFeedback}, atomic spins~\cite{AtomCoolingExper}, and cavity-assisted cooling of mechanical oscillators~\cite{twoCavityFBCool}, among others.

Compared with feedforward interactions where the emitted coherent field or the measurement record is not fed back to the system via a feedback channel, the dynamics of a coherent feedback network is more complex to analyze because of the delay induced by the feedback loop. Different from traditional classical feedback loops, the analysis of the delay in a quantum coherent feedback network is often inadequate if we only consider the wave velocity and the length of the round trip. As already analyzed in Refs.~\cite{delay09,delayPRL}, the existence of the delay not only induces phase shifts among different nodes, but also influences the coherence properties such as correlations by modulating the evolution of quantum states. Hence, delay can work as a control mechanism. For example, when an atomic system is coupled with a waveguide, the evolution of the atomic population is affected by the round trip propagation time of the photon in the waveguide~\cite{photonfeedback}.  When there are two atoms coupled with the waveguide which is ended with a totally reflecting mirror on one side (namely, a semi-infinite waveguide), the dynamics of the two-atom system is affected by the locations of the atoms in the waveguide~\cite{ZhangB}; a similar study can be found in Ref.~\cite{ChiralPRB}.  The steady-state output  two-photon state of a coherent feedback network composed of two atoms and driven by two input photons is derived in Ref.~\cite{zhang2020dynamics}. In addition to loop delay, coupling strength also influences the interaction and exchange of photons among components in a quantum network, and thus affects the performance of the coherent feedback network.

The analysis of coherent feedback with photons becomes even more complicated as the number of excitations increases. Practically, a multi-photon state can be generated with a multi-level atom~\cite{threelevel1998,threeleveltwolevel,pan2017scattering}, a single two-level atom which is repeatedly driven~\cite{twoleveltwophoton}, or multiple two-level atoms~\cite{twoatomtwophoton}. Take the generation and absorption of a two-photon state as an example~\cite{TPabsorption,TPabsorption2,Retardation89,twoleveltwophoton}, which is a generalization of the spontaneous emission of single photons~\cite{Fan09,shen2005coherent,chen2011coherent}, the two-photon interference varies according to the level structure of atoms~\cite{pan2017scattering,dynamicalModel2016}. As theoretically analyzed and experimentally demonstrated in Ref.~\cite{TwophotonNP} on a quantum dot platform, which can be generalized to other platforms such as ion traps~\cite{Iontraptwoatom} and quantum circuits~\cite{lang2013}, the number of emitted photons from a two-level atom is determined by the width of the driving pulse and the two-photon state can be generated by re-exciting the atom right after the first photonic wavepacket is emitted.

In the theoretical and experimental results above-mentioned, waveguide has been commonly used to enable coherent feedback control due to its advantages in the transportation of quantum states~\cite{WQEDtransport} and the preservation of quantum coherence~\cite{WQEDcomputation,RMPWallraff}. For instance, the emitted photon from a cavity coupled with a two-level atom (cavity-QED) can be perfectly reflected  by the end of the waveguide and fed back to the cavity. If the loop delay is properly designed, the photon is able to stabilize the Rabi oscillation in the cavity~\cite{FeedbackRabi}.
Generally speaking, photons transported in the waveguide and reflected by the terminal of the waveguide affect the evolution of the quantum system coupled to the waveguide. The length of the waveguide and the location of the quantum system determine the delay and phase shift induced by the feedback loop, thus regulating the evolution and the steady state of the quantum system. Besides, the effects of coherent feedback are different depending on whether the waveguide and the cavity are coupled through a continuous-mode scheme or a discrete-mode scheme~\cite{photonfeedback,FeedbackRabi}. Specifically, different coherent feedback control mechanisms can be designed according to the size of the waveguide and the coupling methods and strengths to fulfil the required control performance, .g., to maintain the Rabi oscillation~\cite{FeedbackRabi}, generate required atomic or photonic states~\cite{PaolaFeedback,DelayFeedbackPhSq}, or produce entangled quantum states~\cite{ZJL15}.

In most studies of quantum coherent feedback control, it is always assumed that signals' propagation time over the network is small and thus the effect of time delay on the control performance is ignored or treated as a constant phase shift, see for example \cite[Sect. II-B]{Wiseman94}, \cite[Fig. 15]{gough2009series}, \cite[Sect. VI]{GJN10}, and the survey papers \cite{Zhangjing2017,CKS17,ZD22}.
Thus, the system under study is essentially Markovian. If propagation time is treated in a more rigorous way, the system dynamics is non-Markovian and is conceivably more complex. For example, when two or more atoms are coupled with an infinite waveguide~\cite{zhang2020dynamics,Chiralentangle}, photons can bounce back and forth between two atoms, accordingly the system dynamics is complicated by the round-trip transmission delay of the photons between two atoms. Moreover, if a coherent feedback network is closed by a semi-infinite waveguide, photons reflected by the terminal mirror of the waveguide can re-interact with the atom, thus providing another feedback mechanism besides that induced by photons bouncing between two atoms. For this type of coherent feedback networks where a few atoms interacting with a few photons and the whole loop is closed by a semi-infinite waveguide, due to system complexity most existing studies focus on the single-excitation case, see for example~\cite{FeedbackRabi,photonfeedback,ZhangB}.
In this paper, we study the two-excitation case. Specifically, we study how to use coherent feedback to control the atomic evolution and the generation of two-photon states in the coherent feedback network shown in Fig. \ref{fig:scheme}, where the Jaynes-Cummings system \cite{DZWW23} is coupled to a waveguide. Based on the derivation of the relationship between quantum state evolution and the feedback loop length, we propose the optimal parameter design in the continuous coupling scheme to generate two photon states, and illustrate why the two photons cannot be simultaneously observed in the waveguide in the discrete coupling scheme.

The rest of the paper is organized as follows. Section~\ref{Sec:continuous} concentrates on the feedback interaction when the waveguide and the cavity are coupled with continuous modes, especially on the control performance such as the two-photon generation and entanglement influenced by parameter design. The circumstance of the coupling with periodic discrete modes is explored in Section~\ref{Sec:discrete}, which is much different from the continuous coupling scheme studied in Section~\ref{Sec:continuous}. Section~\ref{Sec:conclusion} concludes this paper.

\section{Coherent feedback with the waveguide via continuous coupled modes} \label{Sec:continuous}

\begin{figure}[h]
\centerline{\includegraphics[width=1\columnwidth]{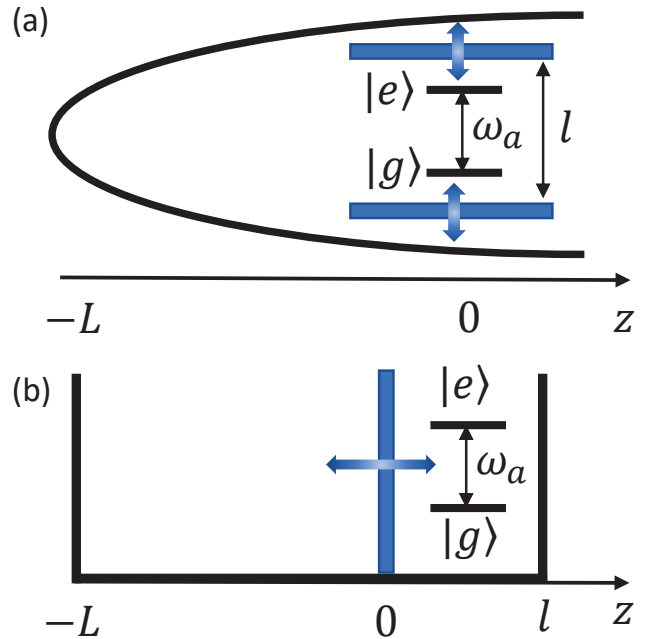}}
\caption{Quantum coherent feedback control of a cavity-QED system coupled to a  waveguide with continuous modes via two semi-transparent mirrors (a) and discrete modes via one semi-transparent mirror (b).}
	\label{fig:scheme}
\end{figure}

We study the quantum coherent feedback network as shown in Fig.~\ref{fig:scheme}. In (a),  the cavity of length $l$ is coupled  with a waveguide of the feedback loop length $2L$ through two semi-transparent mirrors located at $z=0$, as shown by the blue bar. An arbitrary mode in the waveguide can interact with the cavity which contains a two-level atom (represented with the two-way blue arrows). When $L\gg l$,  the cavity and the waveguide are coupled with all the continuous modes of the waveguide~\cite{photonfeedback}. While if the cavity can be only coupled with the reservoir confined at the area $z<0$ through the mirror at $z=0$ as shown in Fig. \ref{fig:scheme}(b), then the waveguide functions similarly as another cavity. When $L$ and $l$ are comparable, the waveguide and cavity are coupled with discrete modes~\cite{CavityEquation,ScullyTreatment,spacetime,Geathreelevel}.

In this section, we consider the circumstance that $L\gg l$ and arbitrary waveguide modes can be coupled with the atomic system independent of the feedback loop length.   The Hamiltonian of the whole quantum system reads
\begin{equation} \label{con:Hamiltonian}
\begin{aligned}
H = H_A + H_{I} + \hbar\omega_c a^{\dag}a +\int \omega_k d_k^{\dag}d_k \mathrm{d}k .
\end{aligned}
\end{equation}
Here, $H_A = \frac{1}{2}\omega_a (|e\rangle \langle e| - |g\rangle \langle g|)$ is the Hamiltonian of the two-level atom with the atomic transition frequency $\omega_a$, $H_{I}$ represents the interaction Hamiltonian between the atom and the cavity as well as that between the cavity and the waveguide. The last two terms of $H$ are the quantized fields in the single-mode cavity and waveguide respectively, where $\omega_c$ is the resonant frequency of the cavity, $\omega_k = ck$ is the frequency of the waveguide mode $k$ with $c$ being the velocity of the coherent fields in the waveguide, $a$($a^{\dag}$) and $d_k$($d^{\dag}_k$) are the annihilation(creation) operators of the cavity and waveguide modes, respectively. In this paper, it is assumed that the atom is resonant with the cavity mode; i.e., $\omega_a = \omega_c$.

\newtheorem{assumption}{Assumption}
\begin{assumption} \label{initial}
Initially, the atom is in the excited state and there is one photon in the cavity.
\end{assumption}
The quantum state of the whole system is of the form:
\begin{equation} \label{con:twophotonState}
\begin{aligned}
&\Psi(t) \\
=&\;  c_e(t) |e,1,\{0\}\rangle + \int_0^\infty c_{ek}(t) |e,0,\{k\}\rangle \mathrm{d}k \\
&+ c_g(t) |g,2,\{0\}\rangle  + \int_0^\infty c_{gk}(t) |g,1,\{k\}\rangle \mathrm{d}k\\
 &+ \int_0^\infty\int_0^{\infty} c_{gkk}(t,k_1,k_2) |g,0,\{k_1\}\{k_2\}\rangle \mathrm{d}k_1\mathrm{d}k_2.
\end{aligned}
\end{equation}
The meaning of each component on the right-hand side  of Eq. \eqref{con:twophotonState} is as follows:\\
~~~~(1)~$|e,1,\{0\}\rangle$: the atom is  in the excited state $\ket{e}$ and there is one photon in the cavity;\\
~~~~(2)~$|e,0,\{k\}\rangle$: the atom is in the excited state and there is one photon in the waveguide with the mode $k$;\\
~~~~(3)~$|g,2,\{0\}\rangle$: the atom is in the ground state $\ket{g}$ and there are two photons in the cavity;\\
~~~~(4)~$|g,1,\{k\}\rangle$: the atom is in the ground state,  one photon is in the cavity and the other is in the waveguide with the mode $k$;\\
~~~~(5)~$|g,0,\{k_1\}\{k_2\}\rangle$: the atom is in the ground state and there are two photons in the waveguide with modes $k_1$ and $k_2$, respectively.

Accordingly, the coefficients (probability amplitudes) $c_e(t)$, $c_g(t)$, $c_{ek}(t)$, $c_{gk}(t)$, $c_{gkk}(t,k_1,k_2)$ represent the time-varying amplitudes of these five states of the quantum system. According to {\bf Assumption~\ref{initial}}, the initial state of the system is  $|\Psi(0)\rangle = |e,1,\{0\}\rangle$. Thus, the initial amplitudes are $c_e(0) =1$, and $c_{ek}(0) = c_g(0) = c_{gk}(0) = c_{gkk}(0) = 0$.

The evolution of the quantum system is governed by the Schr\"{o}dinger equation
\[
\frac{\partial}{\partial t}|\Psi\rangle = -i H |\Psi(t)\rangle,
\]
where the reduced Planck constant $\hbar$ is set to be $1$ throughout this paper. As given in Ref.~\cite{FeedbackRabi}, the interaction Hamiltonian $H_I$ in Eq.  (\ref{con:Hamiltonian}) reads
\begin{equation} \label{con:Ham}
\begin{aligned}
H_I = &- \gamma (\sigma^-a^{\dag} + \sigma^+a)\\
 &-  \int_0^{\infty} \mathrm{d}k\; [G(k,t) a^{\dag}d_k + G^*(k,t)ad^{\dag}_k],
\end{aligned}
\end{equation}
where $\sigma^- = |g\rangle \langle e|$ and $\sigma^+ = |e\rangle \langle g|$ are the lowering and raising operators of the atom respectively, $\gamma$ is the coupling strength between the atom and the cavity, $G(k,t) = G_0\sin(kL)e^{-i(\omega-\Delta_0)t}$ describes the coupling  between the cavity and the waveguide of the mode $k=\frac{\omega}{c}$, where $G_0$ is the magnitude and $\Delta_0 = \omega_a$ is the central mode of the emitted photon~\cite{photonfeedback}.  The theoretical analysis to be conducted in this section is based on the following assumptions which can be fulfilled, as has been discussed in Ref.~\cite{TwophotonNP}.

\begin{assumption} \label{continuous}
The time evolution of the populations are continuous.
\end{assumption}

\begin{assumption} \label{bound}
The photon statistics of infinitely large modes is zero; in other words, $\displaystyle{\lim_{k,k_1,k_2\rightarrow \infty}} (|c_{gk}(t,k)|^2+ |c_{ek}(t,k)|^2 + |c_{gkk}(t,k_1,k_2)|^2) = 0$.
\end{assumption}

Simply speaking, {\bf Assumption} \ref{continuous} is due to the fact that the Schr\"{o}dinger equation generates the unitary evolution of a  quantum system and hence its state evolution   is continuous in time,   while {\bf Assumption} \ref{bound} is due to the fact that atomic emissions produce Lorentzian  photonic pulses. Both {\bf Assumptions} \ref{continuous} and \ref{bound}  are used in the proof of  Lemma \ref{Integr0}.

Substituting Eq. (\ref{con:twophotonState}) into the Schr\"{o}dinger equation yields a system of integro-differential equations
\begin{subequations} \label{con:Popuquation}
\begin{numcases}{}
   \dot{c}_e(t) =  i\sqrt{2}\gamma c_g(t) + i\int_0^\infty c_{ek}(t,k)G(k,t)\mathrm{d}k,  \label{model1}\\
   \dot{c}_{ek}(t,k) = i c_e(t) G^*(k,t)  + i\gamma  c_{gk}(t,k), \label{model2}\\
   \dot{c}_g(t) = i\sqrt{2}\gamma c_e(t) + i\sqrt{2}\int_0^\infty c_{gk}(t,k) G(k,t) \mathrm{d}k,  \label{model3}\\
   \dot{c}_{gk}(t,k) = i\gamma c_{ek}(t,k) + i\sqrt{2}c_g(t) G^*(k,t) \label{model4}\\
 ~~~~~~~~~~~~~~ + i  \int_{0}^{\infty} G(k_2,t) c_{gkk}(t,k,k_2)\mathrm{d}k_2 \notag\\
 ~~~~~~~~~~~~~~ + i  \int_{0}^{\infty} G(k_1,t) c_{gkk}(t,k_1,k)\mathrm{d}k_1, \notag\\
 \dot{c}_{gkk}(t,k_1,k_2) = ic_{gk}(t,k_2)G^*(k_1,t) \label{model5}\\
~~~~~~~~~~~~~~~~~~~ + ic_{gk}(t,k_1)G^*(k_2,t) \notag,
\end{numcases}
\end{subequations}
where the time evolutions of the amplitudes are continuous.
Here, Eq. (\ref{model1}) indicates that tthe state $|e,1,\{0\}\rangle$ can be acquired in two ways: 1) the atom absorbs one photon from the cavity which contains two photons, or 2) the atom is initially excited in an empty cavity and the cavity can further absorb one photon from the waveguide. Eq. (\ref{model2}) indicates that 1) the cavity can emit one photon into the waveguide when the atom is in the excited state and there is one photon in the cavity, or 2) the atom can emit one photon into the cavity. Eq. (\ref{model3}) represents two processes: the spontaneous emission of the excited atom and the  absorption of one photon by the cavity from the waveguide. Eq. (\ref{model4})  shows the exchange of a photon between the waveguide and the cavity. Finally, Eq. (\ref{model5}) means that the waveguide initially having a photon can absorb another one from the cavity to generate a two-photon state.

Generalizing the single-photon feedback scheme studied in Ref.~\cite{photonfeedback}, we are able to derive the control equation for the amplitude of the eigenstate $|e,1,\{0\}\rangle$ for $t\geq 0$ as:
\begin{equation} \label{con:cet7}
\begin{aligned}
\dot{c}_e(t) &=i\sqrt{2}\gamma c_g(t) - \frac{G_0^2\pi}{2c} (c_e(t)-e^{i\Delta_0\tau}c_e(t-\tau)\Theta(t-\tau)),
\end{aligned}
\end{equation}
where $\Theta$ represents the Heaviside step function and $\tau = \frac{2L}{c}$ is the delay induced by the coherent feedback loop. Similarly, the amplitude of the eigenstate $|g,1,\{k\}\rangle$ is:
\begin{equation} \label{con:cgktnewNotau4}
\begin{aligned}
c_{gk}(t,k) =&\; i\sqrt{2}\int_0^t c_g(\nu) G^*(k,\nu)d\nu \\
&- \gamma\int_0^t\int_0^u c_e(\nu) G^*(k,\nu)  d\nu du\\
&-\gamma^2\int_0^t\int_0^u  c_{gk}(\nu,k)d\nu du \\
&-\frac{G_0^2\pi}{c}\int_0^t [c_{gk}(\nu,k)-c_{gk}(\nu-\tau,k)e^{i\Delta_0\tau}]d\nu,  \\
\end{aligned}
\end{equation}
and that for $|g,2,\{0\}\rangle$ is:
\begin{equation} \label{con:cgPopu2}
\begin{aligned}
\dot{c}_g(t) &=i\sqrt{2}\gamma c_e(t) -i\frac{\sqrt{2}\gamma G_0^2\pi}{2c} \tau c_e(t-\tau)e^{i\Delta_0\tau}\Theta(t-\tau)\\
&~~~~- \frac{G_0^2\pi}{c}(c_g(t)-e^{i\Delta_0\tau}c_g(t-\tau)\Theta(t-\tau)).\\
\end{aligned}
\end{equation}
The derivation of Eqs. \eqref{con:cet7}-\eqref{con:cgPopu2} is given in {\bf Appendix~\ref{Sec:Model}}.

Let $\kappa = \frac{\pi G_0^2}{2c}$ denote the coupling strength between the cavity and the waveguide. By means of Eqs. \eqref{con:cet7}-\eqref{con:cgPopu2}, the control equations in Eq.~(\ref{con:Popuquation}) become
\begin{small} 
\begin{subequations}  \label{con:PopuquationResult}
\begin{numcases}{}
\dot{c}_e(t) =  i\sqrt{2}\gamma c_g(t) - \kappa c_e(t) \notag\\
~~~~~~~~~+\kappa e^{i\Delta_0\tau}c_e(t-\tau)\Theta(t-\tau),  \label{controlmodel1}\\
   \dot{c}_{ek}(t,k) = i c_e(t) G^*(k,t)  + i\gamma  c_{gk}(t,k) ,\label{controlmodel2}\\
   \dot{c}_g(t) = i\sqrt{2}\gamma c_e(t) - \kappa c_g(t) + \kappa e^{i\Delta_0\tau}c_g(t-\tau)\Theta(t-\tau) \notag \\
   ~~~~~~~~~ -i\sqrt{2}\gamma\kappa \tau c_e(t-\tau)\Theta(t-\tau)e^{i\Delta_0\tau}, \label{controlmodel3}\\
   \dot{c}_{gk}(t,k) = -  2\kappa c_{gk}(t,k) + 2\kappa c_{gk}(t-\tau,k)\Theta(t-\tau)e^{i\Delta_0\tau} \notag\\
 ~~~~~~~~~~~~~ +  i\sqrt{2}c_g(t) G^*(k,t)-\gamma\int_0^t c_e(t) G^*(k,t) dt \notag\\
 ~~~~~~~~~~~~~ -\int_0^t \gamma^2  c_{gk}(t,k)dt,  \label{controlmodel4}\\
 \dot{c}_{gkk}(t,k_1,k_2) = ic_{gk}(t,k_2)G^*(k_1,t)  \notag\\
 ~~~~~~~~~~~~~~~~~~~~+ ic_{gk}(t,k_1)G^*(k_2,t).  \label{controlmodel5}
\end{numcases}
\end{subequations}
\end{small}

Laplace transforming the amplitudes $c_e(t)$ in Eq. \eqref{controlmodel1} and $c_g(t)$ in Eq. \eqref{controlmodel3} to get their frequency-domain counterparts $C_e(s)$ and $C_g(s)$ of the form
\begin{small}
\begin{equation}\label{eq:c_e_s}
    C_e(s) = \frac{s+\kappa (1-e^{i\Delta_0\tau}e^{-s\tau})}{(s+\kappa (1-e^{i\Delta_0\tau}e^{-s\tau}))^2 + 2\gamma^2 (1-\kappa \tau e^{-s\tau}e^{i\Delta_0\tau})},
\end{equation}
\end{small}
and
\begin{small}
\begin{equation}\label{eq:c_g_s}
C_g(s)=\frac{i\sqrt{2}\gamma (1-\kappa \tau e^{-s\tau}e^{i\Delta_0\tau})}{(s+\kappa (1-e^{i\Delta_0\tau}e^{-s\tau}))^2 + 2\gamma^2 (1-\kappa \tau e^{-s\tau}e^{i\Delta_0\tau})}
\end{equation}
\end{small}
respectively.

In the following subsections, three different scenarios categorized by the length of the feedback loop are studied. In subsection \ref{subsec:small}, the length of the feedback loop is close to zero. In this case, the quantum state oscillates and generates a two-photon state with a small probability.  In subsection \ref{subsec:mid}, the feedback loop is modulated according to the phase shift, and the generated single or two photon states can be controlled by tuning the length of the waveguide. In subsection \ref{subsec:large}, the feedback loop is so long that quantum state evolves within the transmission time of a single round trip in the feedback loop, after that the two photons are both emitted into the waveguide. Finally, in subsection \ref{subsec:entanglement}, two-photon entanglement is analyzed.

\subsection{Feedback control with a waveguide of small length}\label{subsec:small}
When the length of the waveguide $L$ is small,  the induced delay $\tau = \frac{2L}{c}\approx 0$. In this case, $e^{-s\tau} \approx 1$ and
\begin{equation} \label{con:Simple1}
\begin{aligned}
\kappa (1-e^{i\Delta_0\tau}e^{-s\tau}) & \approx  \kappa(1-e^{i\Delta_0\tau}).
\end{aligned}
\end{equation}
Consequently, from Eqs. \eqref{eq:c_e_s} and \eqref{eq:c_g_s} we get
\begin{small}
\begin{eqnarray}
C_e(s)  &\approx&  \frac{s+\kappa(1-e^{i\Delta_0\tau})}{(s+\kappa(1-e^{i\Delta_0\tau}))^2 + 2\gamma^2},
\label{eq:c_e_s_2}
\\
C_g(s) &\approx& \frac{i\sqrt{2}\gamma (1-\kappa \tau e^{i\Delta_0\tau})}{(s+\kappa (1-e^{i\Delta_0\tau}))^2 + 2\gamma^2}.
\label{eq:c_g_s_2}
\end{eqnarray}
\end{small}
Eqs. \eqref{eq:c_e_s_2}, \eqref{eq:c_g_s_2} and  \eqref{controlmodel4} yield
\begin{small}
\begin{equation} \label{con:CgkddotApproxi}
\begin{aligned}
&\ddot{c}_{gk}(t,k)
\\
&\approx -  2\kappa[\dot{c}_{gk}(t,k)-\dot{c}_{gk}(t-\tau,k)e^{i\Delta_0\tau}] - \gamma^2  c_{gk}(t,k)\\
&~~~~+G_0\sin(kL) \{(D-\frac{3\gamma}{2})e^{[E+i(\omega-\Delta_0 + F +\sqrt{2}\gamma)]t} \\
&~~~~-(D+\frac{3\gamma}{2}) e^{[E+i(\omega-\Delta_0+F -\sqrt{2}\gamma)]t}\},
\end{aligned}
\end{equation}
\end{small}
where
\begin{small}
\begin{equation} \label{con:jan8_1}
\begin{aligned}
&D = \frac{\sqrt{2}}{2} [\Delta_0-\omega-\sin(\Delta_0\tau) + i\kappa(\cos(\Delta_0\tau) -1)],
\\
& E = \kappa (\cos\Delta_0\tau -1),
\\
&F = \sin\Delta_0\tau.
\end{aligned}
\end{equation}
\end{small}
Denote
\begin{equation}\label{con:jan8_2}
R =\kappa(e^{i\Delta_0\tau}-1).
\end{equation}
Then the Laplace transform of $c_{gk}(t,k)$ w.r.t. $t$  gives the frequency-domain function $C_{gk}(s,k)$ of the from
\begin{small}
\begin{equation*}
\begin{aligned}
  & C_{gk}(s,k) \\
  =&\; -3G_0\sin(kL)
  \\
  &\; \times \gamma\Bigg[\frac{H(\omega)}{s-R+\sqrt{R^2-\gamma^2}} + \frac{I(\omega)}{s-R -\sqrt{R^2 -\gamma^2}}
  \\
  &\; + \frac{J(\omega)}{s-E-i(\omega-\Delta_0+F) + i\sqrt{2}\gamma}
  \\
  &\; + \frac{K(\omega)}{s-E-i(\omega-\Delta_0+F) - i\sqrt{2}\gamma}\Bigg],
\end{aligned}
\end{equation*}
\end{small}
 where the parameters $H(\omega)$, $I(\omega)$, $J(\omega)$ and $K(\omega)$ are given in {\bf Appendix~\ref{Sec:proofLemma1}}.

Applying the inverse Laplace transform to $C_e(s)$, $C_g(s)$ and $C_{gk}(s,k)$ obtained above gives $c_e(t)$, $c_g(t)$ and $c_{gk}(t,k)$ in the time domain, which are
\begin{small}
\begin{equation} \label{con:CeCglaplace}
\left\{
\begin{aligned}
c_e(t) &=  e^{-\kappa[1-\cos(\Delta_0\tau)]t} e^{i\sin(\Delta_0\tau)t}\cos(\sqrt{2}\gamma t),\\
c_g(t) & = i e^{-\kappa(1-\cos(\Delta_0\tau))t}e^{i\sin(\Delta_0\tau)t}\sin(\sqrt{2}\gamma t),\\
c_{gk}(t,k) &\approx -3G_0\sin(kL) \gamma \{H(\omega)e^{-i\gamma t} + I(\omega)e^{i\gamma t} \\
&~~~~ + J(\omega) e^{[E+i(\omega-\Delta_0+F) - i\sqrt{2}\gamma]t}\\
&~~~~ + K(\omega)e^{[E+i(\omega-\Delta_0+F) + i\sqrt{2}\gamma]t}\}
\end{aligned}
\right.
\end{equation}
\end{small}
respectively.

Before presenting the main result in this subsection, we state the following lemma first.
\newtheorem{lemma}{Lemma}
\begin{lemma}\label{tinytau}
When $\kappa \tau\ll 1$, for the  mode $k$ with frequency $\omega = kc\in[0,2\Delta_0]$,  we have $H(\omega) =I(2\Delta_0-\omega)^*$, and $J(\omega) =K(2\Delta_0-\omega)^*$, where ``$*$'' represents the complex conjugate.
\end{lemma}

\newtheorem*{Proof}{Proof}
\begin{Proof}
See {\bf Appendix~\ref{Sec:proofLemma1}}. \qed
\end{Proof}

The following is the main result of this subsection.
\newtheorem{Theorem}{Theorem}
\begin{Theorem} \label{pureimage}
When $\kappa \tau \ll 1$ and $k_1+k_2 = \frac{2\Delta_0}{c}$, the amplitude $c_{gkk}$ of the two-photon state in the waveguide is purely imaginary.
\end{Theorem}

\begin{Proof}
Consider  $\dot{c}_{gkk}(t,k_1,k_2)$ in Eq.~(\ref{model5}) and denote $\omega_1 = k_1c$ and $\omega_2 = k_2c$. When $\kappa\tau \ll 1$, we have $F\approx0$ in Eq. \eqref{con:jan8_1}. Hence,
\begin{small}
\begin{equation} \label{con:CgkkdotPart1}
\begin{aligned}
&~~~~ic_{gk}(t,k_2)G^*(k_1,t) \\
&= -3iG_0^2\sin(k_2L) \sin(k_1L) \gamma \{H(\omega_2)e^{i(\omega_1 -\Delta_0 -\gamma) t}\\
&~~~~ + I(\omega_2)e^{i(\omega_1 -\Delta_0 +\gamma) t} + J(\omega_2) e^{[E+i(\omega_1+\omega_2-2\Delta_0+F) - i\sqrt{2}\gamma]t}\\
&~~~~ + K(\omega_2)e^{[E+i(\omega_1+\omega_2-2\Delta_0+F) + i\sqrt{2}\gamma]t}\}\\
 &\approx  -3iG_0^2\sin(k_2L) \sin(k_1L) \gamma \{H(\omega_2)e^{i(\omega_1 -\Delta_0 -\gamma) t}\\
&~~~~  + I(\omega_2)e^{i(\omega_1 -\Delta_0 +\gamma) t} + J(\omega_2) e^{[E+i(\omega_1+\omega_2-2\Delta_0) - i\sqrt{2}\gamma]t}\\
&~~~~ + K(\omega_2)e^{[E+i(\omega_1+\omega_2-2\Delta_0) + i\sqrt{2}\gamma]t}\}.
\end{aligned}
\end{equation}
\end{small}
 When $\omega_1 + \omega_2= 2\Delta_0$, 
\begin{small}
\begin{equation} \label{con:CgkkdotPart1Special}
\begin{aligned}
&~~~~ic_{gk}(t,k_2)G^*(k_1,t) \\
&\approx  -3iG_0^2\sin(k_2L) \sin(k_1L) \gamma \{H(\omega_2)e^{i(\omega_1 -\Delta_0 -\gamma) t}\\
&~~~~ + I(\omega_2)e^{i(\omega_1 -\Delta_0 +\gamma) t}+ J(\omega_2) e^{(E - i\sqrt{2}\gamma)t}\\
&~~~~ + K(\omega_2)e^{(E + i\sqrt{2}\gamma)t}\}.
\end{aligned}
\end{equation}
\end{small}
According to {\bf Lemma~\ref{tinytau}},
\begin{small}
\begin{equation} \label{con:Property}
\left\{
\begin{aligned}
&H(\omega_2) =I(\omega_1)^*,\\
&\omega_1 -\Delta_0-\gamma = -(\omega_2 -\Delta_0 +\gamma), \\
&J(\omega_2) =K(\omega_1)^*.
\end{aligned}
\right.
\end{equation}
\end{small}
Therefore,
\begin{small}
\begin{equation} \label{con:PropertyAdd}
\left\{
\begin{aligned}
&H(\omega_2)e^{i(\omega_1 -\Delta_0 -\gamma) t} + I(\omega_1)e^{i(\omega_2 -\Delta_0 +\gamma) t} \\
=&\; 2\Re(H(\omega_2)e^{i(\omega_1 -\Delta_0 -\gamma) t}) ,\\
&H(\omega_1)e^{i(\omega_2 -\Delta_0 -\gamma) t} + I(\omega_2)e^{i(\omega_1 -\Delta_0 +\gamma) t}\\
=&\; 2\Re(H(\omega_1)e^{i(\omega_2 -\Delta_0 -\gamma) t}) ,\\
&J(\omega_2) e^{(E - i\sqrt{2}\gamma)t} + K(\omega_1) e^{(E + i\sqrt{2}\gamma)t} \\
=&\;2\Re(K(\omega_1) e^{(E + i\sqrt{2}\gamma)t}) ,\\
&K(\omega_2)e^{(E + i\sqrt{2}\gamma)t} + J(\omega_1)e^{(E - i\sqrt{2}\gamma)t}\\
=&\; 2\Re(K(\omega_2) e^{(E + i\sqrt{2}\gamma)t}),
\end{aligned}
\right.
\end{equation}
\end{small}
where $\Re$ represents the real part of a complex number. Thus, when $\omega_1+\omega_2 = 2\Delta_0$,
\begin{small}
\begin{equation} \label{con:CgkkdotConclusion}
\begin{aligned}
&\dot{c}_{gkk}(t,k_1,k_2) \approx  -6iG_0^2\sin(k_2L) \sin(k_1L) \gamma\\
&~~~~ [\Re(H(\omega_2)e^{i(\omega_1 -\Delta_0 -\gamma) t}) + \Re(H(\omega_1)e^{i(\omega_2 -\Delta_0 -\gamma) t})\\
 &~~~~+ \Re(K(\omega_1) e^{(E + i\sqrt{2}\gamma)t}) + \Re(K(\omega_2) e^{(E + i\sqrt{2}\gamma)t})],
\end{aligned}
\end{equation}
\end{small}
which is a purely imaginary number. Because the initial condition is $c_{gkk}(0,k_1,k_2) = 0$ as given in {\bf Assumption~\ref{initial}},  $c_{gkk}(t,k_1,k_2)$ is also a purely imaginary number for all $t\geq0$. \qed
\end{Proof}

\begin{figure}[h]
\centerline{\includegraphics[width=1.1\columnwidth]{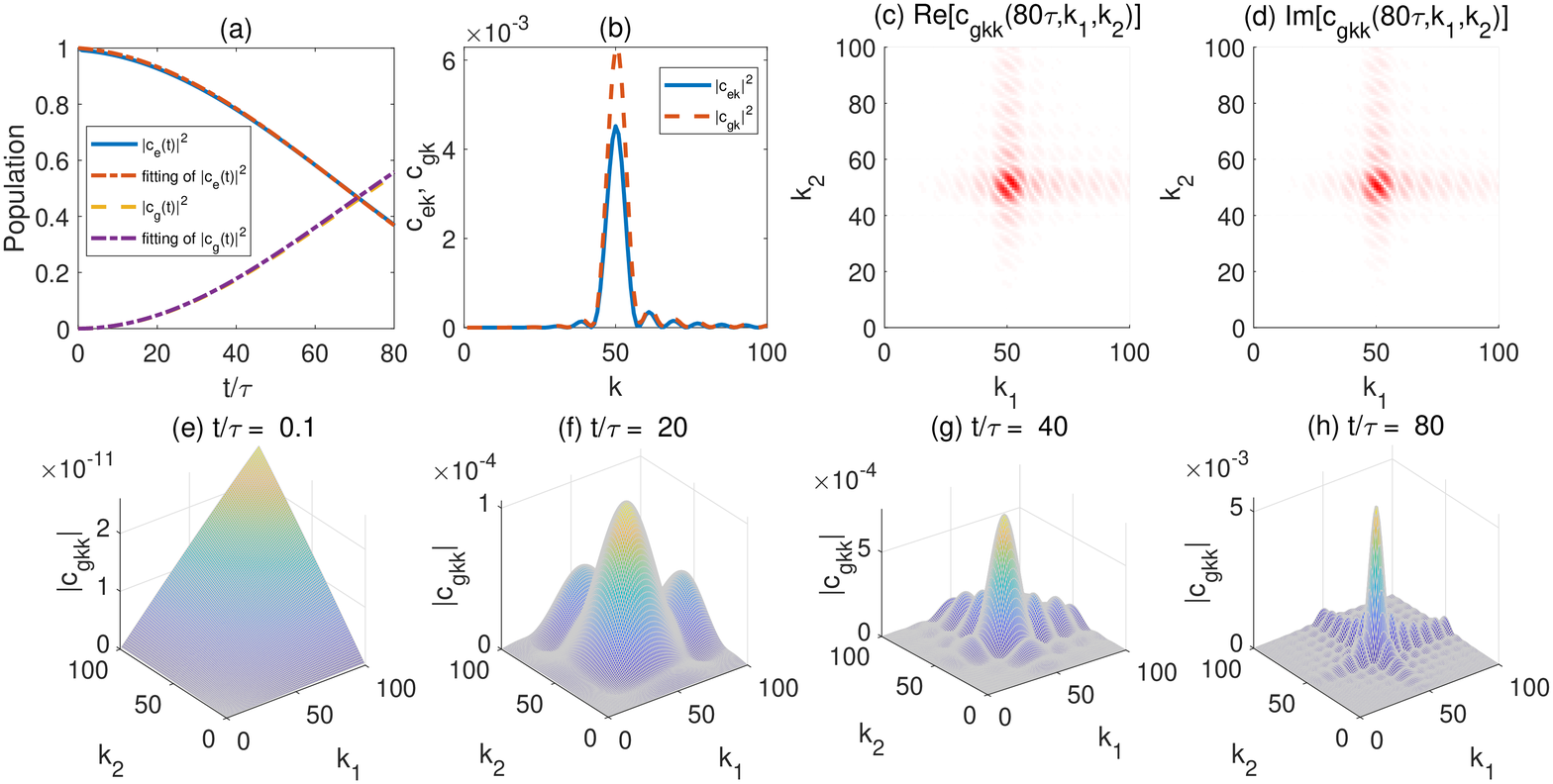}}
\caption{Simulations of the coherent feedback with short waveguide $L = 0.005$m and cavity-waveguide coupling $G_0 = 0.5$.}
	\label{fig:shortwaveguide}
\end{figure}

The numerical simulations are shown in Fig. \ref{fig:shortwaveguide}, where $k_{1,2}\in[0,100]$, $\Delta_0 = 50$ and $\gamma = 2\kappa$. As the length of the waveguide $L$ as well as the coupling between the waveguide and the cavity $\kappa$ (defined before Eq.~(\ref{con:PopuquationResult})) is small, it can be seen from Fig. \ref{fig:shortwaveguide}(a)  that the atom  oscillates between its excited state and ground state, where $|c_e(t)|^2$ and $|c_g(t)|^2$ are the numerical simulations with Eq.~(\ref{con:PopuquationResult}) and the dash-dot lines represent the fitting results of the populations with Eqs. \eqref{eq:c_e_s_2}-\eqref{eq:c_g_s_2} based on the approximation in Eq.~(\ref{con:Simple1}). Therefore, the populations of the single-photon state in Fig. \ref{fig:shortwaveguide}(b) and the generated two-photon state in Fig. \ref{fig:shortwaveguide}(e)-(h) are much smaller than the oscillation amplitude in Fig. \ref{fig:shortwaveguide}(a). Fig. \ref{fig:shortwaveguide}(c)-(d) show  the real and imaginary parts of $c_{gkk}$ when $t= 80\tau$, which agree with the conclusion in {\bf Theorem~\ref{pureimage}}.

\newtheorem{remark}{Remark}
\begin{remark}
Recall that $\gamma$ is the coupling strength between the atom and the cavity.  If the atom is initialized in the excited state, a large $\gamma$ will enhance the efficiency of spontaneously emitting a photon into the cavity. This, together with a large $G_0$, will further induce a big probability of detecting two photons in the waveguide. This agrees with Eq. (\ref{con:CgkkdotConclusion}) which says that large $\gamma$ and $G_0$ induce large amplitudes of $\dot{c}_{gkk}$.
\end{remark}

\begin{remark}
As $\Delta_0$ is the central mode of the continuous-mode photonic field, which should be theoretically much larger than the scale of $\tau$. In the numerical simulations, $\Delta_0 = 50$, which is large enough to ensure that $\Delta_0\tau \gg \tau$. As a result, the approximation in Eq. (\ref{con:Simple1}) is much easier to be satisfied than that in {\bf Lemma~\ref{tinytau}}. Additionally, a large range of $\Delta_0\tau$ provides an approach to controlling the feedback dynamics by tuning the length of the waveguide to get the suitable $\tau$.
\end{remark}

\subsection{Feedback control with a waveguide of finite length}\label{subsec:mid}

In the case of coherent feedback control via a waveguide of finite length, we have the following result.

\begin{Theorem}\label{thm:2}
When $\Delta_0\tau = n\pi$ and $\tau \ll 1$, $n = 1,2,\cdots$, the two-photon amplitude $c_{gkk}$ is purely imaginary provided that $\omega_1+\omega_2 = 2\Delta_0$.
\end{Theorem}

\begin{Proof}
1) When $n$ is an even number, $\cos(\Delta_0\tau) = 1$ and $\sin(\Delta_0\tau) = 0$. The calculation is the same as the proof of {\bf Theorem~\ref{pureimage}}. \\
2) When $n$ is an odd number, $\cos(\Delta_0\tau) = -1$, and  $\sin(\Delta_0\tau) = 0$. By Eqs. \eqref{con:jan8_1} and \eqref{con:jan8_2}, $R =\kappa(e^{i\Delta_0\tau}-1) = -2\kappa$, $E = \kappa (\cos\Delta_0\tau -1) = -2\kappa$, $F = \sin\Delta_0\tau = 0$, and $D = \frac{\sqrt{2}}{2} [\Delta_0-\omega-\sin(\Delta_0\tau) + i\kappa(\cos(\Delta_0\tau) -1)] = \frac{\sqrt{2}}{2} [\Delta_0-\omega -2i\kappa]$. Denote $M = \omega-\Delta_0+F$. Then
\begin{small}
\begin{equation} \label{con:CgkddotMiddletauMN}
\begin{aligned}
&C_{gk}(s,k) \\
=&\; G_0\sin(kL)\frac{2\sqrt{2}i\gamma D -3\gamma [s-E-iM]}{(s^2 - 2R s + \gamma^2)[(s-E-iM)^2 +2\gamma^2]}\\
=&\; G_0\sin(kL) \Bigg[\frac{H(\omega)}{s-R+\sqrt{R^2-\gamma^2}} + \frac{I(\omega)}{s-R-\sqrt{R^2-\gamma^2}}\\
& + \frac{J(\omega)}{s-E-iM + i\sqrt{2}\gamma} + \frac{K(\omega)}{s-E-iM - i\sqrt{2}\gamma}\Bigg].
\end{aligned}
\end{equation}
\end{small}
Moreover, it is easy to show that Eq. (\ref{con:omega12}) still holds. As a result, by Eq. (\ref{con:CgkkdotConclusion}) in the proof of {\bf Theorem~\ref{pureimage}}, $c_{gkk}$ is purely imaginary. \qed
\end{Proof}

Physically, Theorem \ref{thm:2} means that when $\Delta_0\tau = n\pi$ and $\omega_1+\omega_2 = 2\Delta_0$,  the amplitude of the two-photon state component of the whole state $\Psi(t)$ in Eq.~\eqref{con:twophotonState} is a pure phase.

\begin{Theorem}\label{npi}
When $\Delta_0\tau = 2n\pi$, $n = 1,2,\cdots$, the single-photon state $|g,1,\{k\}\rangle$ oscillates and does not decay to zero. When $\Delta_0\tau \neq 2n\pi$, eventually there are two photons in the waveguide.
\end{Theorem}

\begin{figure}[h]
\centerline{\includegraphics[width=0.95\columnwidth]{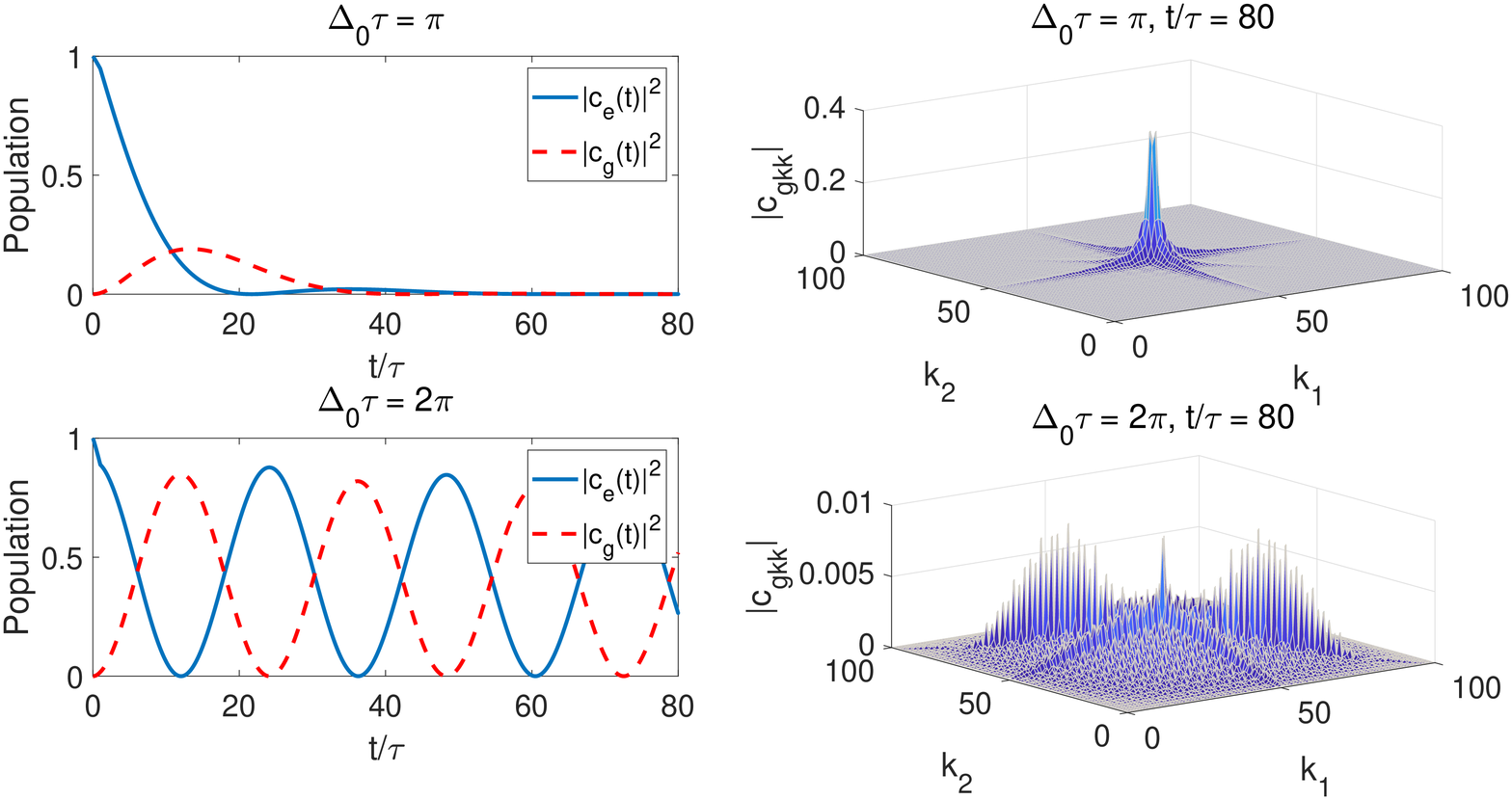}}
\caption{Coherent feedback control with waveguides of different length (upper:  $\Delta_0\tau = \pi$ and $L = 0.0314$m; lower: $\Delta_0\tau  = 2\pi$ and $L =0.0628$m).}
	\label{fig:population}
\end{figure}

\begin{Proof}
1). When $\Delta_0\tau = 2n\pi$, $E = \kappa (\cos\Delta_0\tau -1)= 0$, and $R =\kappa(e^{i\Delta_0\tau}-1) = 0$.  In this case, by Eq. (\ref{con:CeCglaplace}), the amplitude of the single-photon state $|g,1,\{k\}\rangle$ is
\begin{small}
\begin{equation} \label{con:Cgksolutiontaull1}
\begin{aligned}
c_{gk}(t,k) &\approx -3G_0\sin(kL) \gamma \{H(k)e^{-i\gamma t}\\
 &+ I(k)e^{i\gamma t} + J(k) e^{[E+i(\omega-\Delta_0+F) - i\sqrt{2}\gamma]t}\\
 &+ K(k)e^{[E+i(\omega-\Delta_0+F) + i\sqrt{2}\gamma]t}\}. \\
\end{aligned}
\end{equation}
\end{small}
Clearly, the four components of $c_{gk}(t,k)$ oscillate and $\lim_{t\rightarrow \infty} c_{gk}(t,k) \neq 0$.\\
2). When $\Delta_0\tau \neq 2n\pi$, 
$\Re (E) < 0$, and $\Re (R) < 0$. By Eq. \eqref{con:CgkddotMiddletauMN},  $\lim_{t\rightarrow \infty} c_{gk}(t,k) = 0$. Then, by Eq. (\ref{con:PopuquationResult}) and its Laplace transform, we get $\lim_{t\rightarrow \infty} c_{e}(t) = \lim_{t\rightarrow \infty} c_{e}(t) = 0$. Moreover, by Eq. (\ref{con:Popuquation}), \\  $\lim_{t\rightarrow \infty}c_{ek}(t,k) = \lim_{t\rightarrow \infty} \dot{c}_{gkk}(t,k_1,k_2) = 0$. Consequently,  the dominant population of the steady state of the quantum system is $|c_{gkk}|^2$. \qed
\end{Proof}
\begin{remark} \label{lenghtcontrol}
When $\Delta_0\tau = (2n-1)\pi$, $E = R =-2\kappa$. In this case, $c_{gk}(t,k)$ decays to zero at the fastest speed for given $\kappa$ and $\gamma$. This means that the two-photon state is most easily generated when $\Delta_0\tau = (2n-1)\pi$.
Therefore, the population terms in the state as well as the number of photons in the waveguide can be controlled by tuning the length of the waveguide. For example, if  the length of the waveguide  $L$ is chosen to be close to the discrete periodic sequence $\frac{n\pi c}{\Delta_0}$, then from the proof of {\bf Theorem \ref{npi}}, it is easy to see that two-photon states are difficult to generate. On the other hand, if we choose  $L = \frac{(2n-1)\pi c}{2\Delta_0}$, then due to $\Delta_0 \gg 1$, the two-photon state can be generated even with a short waveguide. These  agree with the comparisons of the simulations in Fig. \ref{fig:population}.
\end{remark}

The numerical simulations are shown in Fig. \ref{fig:population}, where $\omega_{1,2}\in[0,100]$, $\Delta_0 = 50$, $G_0$ = 0.5, $\gamma = 2\kappa$, $\Delta_0\tau = \pi,2\pi$, respectively. When $\Delta_0\tau = \pi$ as shown in the upper two sub-figures of Fig. \ref{fig:population}, $E = \Re (R) < 0$ in Eq. (\ref{con:CgkddotMiddletauMN}), then $\lim_{t\rightarrow \infty} c_e(t) = \lim_{t\rightarrow \infty} c_g(t) = \lim_{t\rightarrow \infty} c_{gk}(t) = 0$, and $\dot{c}_e(t) \approx 0$ according to Eq. (\ref{controlmodel1}). Therefore, $c_{ek}(t,k) \approx 0$. This means there will be two photons in the waveguide, as indicated by the peak of the amplitude $|c_{gkk}|$ when $t=80\tau$. However, the results are different at $\Delta_0\tau = 2\pi$ as shown in the lower two sub-figures of Fig. \ref{fig:population}, where both $|c_e(t)|^2$ and $|c_g(t)|^2$ oscillate persistently and the amplitude of the two-photon state $c_{gkk}$ is close to zero. The comparison of the simulation results in Fig. \ref{fig:population} confirms {\bf Theorem~\ref{npi}} and {\bf Remark~\ref{lenghtcontrol}}.

\subsection{Feedback control with a long waveguide} \label{subsec:large}

When the coherent feedback loop is long enough that the evolution time of the quantum system is shorter than the transmission delay from the cavity to the terminal end of the waveguide, 
$\delta(t-t'-\tau) =\delta(t-t'+\tau) = 0$ in Eq. (\ref{con:cet2}). In this case, by Eqs. (\ref{con:cet7}) and (\ref{con:cgPopu2_new}), when  $\tau \rightarrow \infty$ the equation of the amplitude with no photons in the waveguide is
\begin{equation} \label{con:ddcet7LL2}
\begin{aligned}
\ddot{c}_*(t) + 3\kappa\dot{c}_*(t) + [2\gamma^2 + 2\kappa^2]c_*(t) &= 0,
\end{aligned}
\end{equation}
where $c_*(t) = c_e(t)$ and $c_g(t)$, respectively.

The following is the main result of this subsection.
\begin{Theorem} \label{thm:no feedback}
When the coherent feedback loop is long enough and the cavity-QED system's evolution time is much shorter than the loop delay, there are two photons in the waveguide when $t\rightarrow \frac{L}{c}$, and the two-photon emission rate is maximized when $\kappa > 2\sqrt{2}\gamma$.
\end{Theorem}
\begin{Proof} Because $\kappa > 0$, by Eq. \eqref{con:ddcet7LL2} we have  $\lim_{t\rightarrow\infty} c_e(t) = \lim_{t\rightarrow\infty} c_g(t) = 0$.
Denote $\Omega_0 = \sqrt{|(\frac{\kappa}{2})^2 - 2\gamma^2|}$.
When $\kappa > 2\sqrt{2}\gamma$, $0<\Omega_0 < \frac{\kappa}{2}$. Solving Eq. \eqref{con:ddcet7LL2} we get $c_*(t) = A_* e^{(-\frac{3}{2}\kappa + \Omega_0)t} + B_*  e^{(-\frac{3}{2}\kappa -\Omega_0)t}$, which decays to zero without oscillations. On the other hand, when $\kappa = 2\sqrt{2}\gamma$, $c_*(t) = (A_*+B_*t) e^{-\frac{3}{2}\kappa t}$; and when  $\kappa < 2\sqrt{2}\gamma$, $c_*(t) = A_* e^{(-\frac{3}{2}\kappa + i \Omega_0)t} + B_*  e^{(-\frac{3}{2}\kappa -i\Omega_0)t}$. In both of these two cases there are oscillations in the evolution of $c_*(t)$.

Denote $\tilde{p}(t,k) = c_e(t)G^*(t,k)$, $\tilde{q}(k,t) = c_g(t)G^*(t,k)$, and take $\Theta(t-\tau) = 0$ in Eq.~(\ref{controlmodel4}). Then we have
\[
\lim_{s\rightarrow 0}sC_{gk}(s,k) = \lim_{s\rightarrow 0}s\frac{i\sqrt{2}s\tilde{Q}(s,k)-\gamma \tilde{P}(s,k)}{s^2+2\kappa s +\gamma^2}=0,
\]
where $\tilde{Q}(s,k)$ and $\tilde{P}(s,k)$ are the Laplace transform of $\tilde{p}(t,k)$ and $\tilde{q}(t,k)$, respectively. Then $\lim_{t\rightarrow \infty}c_{gk}(t,k) = 0$, and $\lim_{t\rightarrow \infty}c_{ek}(t,k) = 0$ according to Eq.~(\ref{controlmodel2}). As the populations of one-photon states in the waveguide are close to zero, eventually there are two photons in the waveguide. \qed
\end{Proof}

\begin{figure}[h]
\centerline{\includegraphics[width=1\columnwidth]{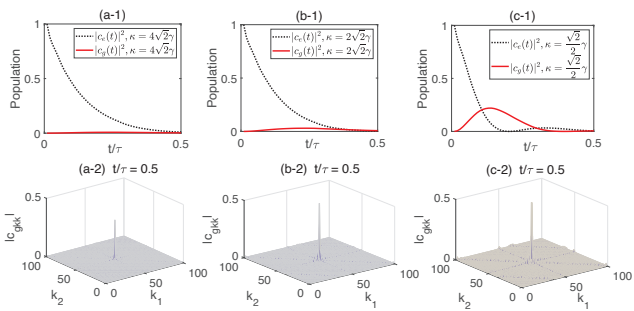}}  
\caption{Coherent feedback control with a long waveguide.}
	\label{fig:LongWaveguideCom}
\end{figure}

In Fig. \ref{fig:LongWaveguideCom}, we take $L =5m$, which is around $80$ times larger than the  simulation with $L=0.0628m$ in the bottom-right subfigure of Fig.~\ref{fig:population}, $\Delta_0 = 50$, $G_0 = 0.5$, and compare these three circumstances in {\bf Theorem \ref{thm:no feedback}} by setting $\kappa = 4\sqrt{2}\gamma$, $\kappa = 2\sqrt{2}\gamma$, and $\kappa = \frac{\sqrt{2}}{2}\gamma$, respectively. The simulations indicate that the populations $|c_e(t)|^2$ and $|c_g(t)|^2$ oscillate when $\kappa \leq 2\sqrt{2}\gamma$, while finally there are two photons in the waveguide in all the parameter settings of $\kappa$ and $\gamma$ if only the waveguide is long enough, and this agrees with the conclusion in {\bf Theorem \ref{thm:no feedback}}.

\subsection{Two-photon entanglement}\label{subsec:entanglement}

In this subsection, we study entanglement property of the two photons in the waveguide.

According to \cite{LiaoEntangle},  \cite{Z14}, \cite{Z17}, \cite{pan2017scattering}, \cite{DZA19a}, \cite{DZA19b}, a two-photon state is an entangled state if the amplitude $c_{gkk}(k_1,k_2)$ in Eq. (\ref{con:twophotonState}) cannot be fatorized as a product of two functions, one with argument $k_1$ and the other with argument $k_2$.

For the coherent feedback scheme in Fig. \ref{fig:scheme}(a), according to {\bf Theorem~\ref{npi}} and {\bf Remark~\ref{lenghtcontrol}}, the most efficient way for the generation of two-photon states occurs when $\kappa\tau = (2n+1)\pi$. In this case, the steady-state amplitude is 
\begin{scriptsize}
\begin{equation} \label{con:CgkkinfiniteDeltak}
\begin{aligned}
&c_{gkk}(\infty,k_1,k_2)\\
=&\;\lim_{t\rightarrow\infty}c_{gkk}(t,k_1,k_2)\\
=&\; -i G_0^2\sin(k_1L)\sin(k_2L)\\
& \Bigg[\frac{H(k_2)}{R-\sqrt{R^2 -\gamma^2} + i\delta(k_1)} +\frac{I(k_2)}{R+\sqrt{R^2 -\gamma^2} + i\delta(k_1)} \\
&+\frac{H(k_1)}{R-\sqrt{R^2 -\gamma^2} + i\delta(k_2)} +\frac{I(k_1)}{R+\sqrt{R^2 -\gamma^2} + i\delta(k_2)} \\
&+ \frac{J(k_2)}{R + i(\delta(k_1) + \delta(k_2)-\sqrt{2}\gamma) } +  \frac{K(k_2)}{R + i(\delta(k_1) + \delta(k_2)+\sqrt{2}\gamma) }\\
&+ \frac{J(k_1)}{R + i(\delta(k_1) + \delta(k_2)-\sqrt{2}\gamma)} +  \frac{K(k_1)}{R + i(\delta(k_1) +\delta(k_2)+\sqrt{2}\gamma)}\Bigg],
\end{aligned}
\end{equation}
\end{scriptsize}
where $R = -2\kappa$, $\delta(k_1) =ck_1 -\Delta_0 $, $\delta(k_2) =ck_2 -\Delta_0 $,  and the other parameters $H,I,J,K$ can be calculated according to Eq. (\ref{con:HIJK}). Obviously, $c_{gkk}(\infty,k_1,k_2)$ is not factorizable in terms of $k_1$ and $k_2$, thus the generated two photons by the coherent feedback are entangled. The conclusion also applies to the circumstance that the waveguide is infinitely long because the factorizability will not be changed with the rising of the feedback loop length. Finally, the degree of entanglement is largely determined by  the poles of $c_{gkk}(\infty,k_1,k_2)$ in Eq. (\ref{con:CgkkinfiniteDeltak}). As the last four terms of the right-hand side of Eq.  (\ref{con:CgkkinfiniteDeltak}) indicate that the non-factorizable  $c_{gkk}(t,k_1,k_2)$ is dominant when $\kappa$ is close to zero,  the generated two photons are more easily entangled when $\kappa$ is sufficiently small.

\begin{remark} \label{entanglement}
 Practically, when the coupling between the waveguide and the cavity is small, the two correlated photons are slowly emitted into the waveguide. This also agrees with the simulation results in Fig.\ref{fig:LongWaveguideCom} where $|c_g(t)|^2$ oscillates more significantly for smaller $\kappa$.

\end{remark}

\section{Quantum feedback control with discrete modes} \label{Sec:discrete}
The discrete coupling between the cavity and the waveguide can be realized by tuning the length $l$ of the cavity and $L$ of the waveguide in Fig. \ref{fig:scheme}(b). Thus, the setting $l\ll L$ used in Section \ref{Sec:continuous}  need not necessarily hold in the discrete coupling scheme. In this section, we study the feedback control of the system in Fig. \ref{fig:scheme}(b)  with discrete coupling modes
as discussed in  \cite{photonfeedback}, and the coupling strength is
\begin{equation} \label{con:discreteCoupling}
\begin{aligned}
G_q(t) \equiv G(k_q,t) = G_0\sin(k_qL)e^{-i(\omega_q-\Delta_0)t},
\end{aligned}
\end{equation}
where $k_q = \frac{(2q+1)\pi}{2L}$ and $q = 0,\pm1,\pm2,\ldots$. This parameter design of the coupling between the cavity and waveguide is in the most efficient fashion because the amplitude of the cavity field induced by the waveguide field via the coherent feedback is maximized, as theoretically analyzed in Ref.~\cite{CavityEquation}; see also {\bf Appendix~\ref{Sec:Maxwell}}. This scheme has been widely used in the analysis of the quasimodes of the cavity coupled with various systems, see, e.g., Refs.~\cite{CavityEquation,ScullyTreatment,spacetime,Geathreelevel}.

The interaction Hamiltonian reads
\begin{small}
\begin{equation} \label{con:discreteHamiltonian}
\begin{aligned}
H_I = & - \gamma (\sigma^-a^{\dag} + \sigma^+a) \\
&- \sqrt{\frac{\pi}{2L}}\sum_{q=-\infty}^{\infty} ( G_q(t)a^{\dag}d_q +  G_q^*(t)d_q^{\dag}a),
\end{aligned}
\end{equation}
\end{small}
where ``$\dagger$'' means the adjoint of an operator while ``$\ast$'' means the complex conjugate of a complex number.

In contrast to Eq. (\ref{con:twophotonState}), the overall system state under the discrete coupling is:
\begin{small}
\begin{equation} \label{con:twophotonStateDiscrete}
\begin{aligned}
\Psi(t) = &c_e(t) |e,1,\{0\}\rangle + \sum_{q=-\infty}^{\infty} c_{eq}(t) |e,0,\{k_q\}\rangle \\
& + c_g(t) |g,2,\{0\}\rangle  + \sum_{q=-\infty}^{\infty} c_{gq}(t) |g,1,\{k_q\}\rangle\\
 &+\sum_{p,q=-\infty}^{\infty} c_{gpq}(t,k_p,k_q) |g,0,\{k_p\}\{k_q\}\rangle \mathrm{d}k_p\mathrm{d}k_q,
\end{aligned}
\end{equation}
\end{small}
where $c_e(t)$ is the amplitude that the atom is in the excited state and there is one photon in the cavity,  $c_g(t)$ is the amplitude that the atom is in the ground state and there are two photons in the cavity, $c_{eq}(t)$, $c_{gq}(t)$ and $c_{gpq}(t,k_p,k_q)$ represent the amplitudes with the discrete modes of photons with modes $k_q$ and $k_p$, respectively.

In analogy to Eq. (\ref{con:Popuquation}), the system of control equations with the delayed feedback loop is
\begin{scriptsize}
\begin{subequations} \label{con:PopuquationResultdiscrete}
\begin{numcases}{}
\dot{c}_e(t) =  i\sqrt{2}\gamma c_g(t)  \label{discretemodel1}\\
 ~~~~~~~~+ i\sqrt{\frac{\pi}{2L}} G_0 \sum_{q= -\infty}^{\infty} c_{eq}(t,k_q) (-1)^q e^{-i(\omega_q-\Delta_0)t},    \notag\\
\dot{c}_{eq}(t,k_q) = i \sqrt{\frac{\pi}{2L}} G_0 (-1)^q c_e(t) e^{i(\omega_q-\Delta_0)t}   \label{discretemodel2}\\
 ~~~~~~~~ +i\gamma  c_{gq}(t,k_q), \notag\\
\dot{c}_g(t) = i\sqrt{2}\gamma c_e(t)   \label{discretemodel3} \\
  ~~~~~~~+ i\sqrt{2} \sqrt{\frac{\pi}{2L}} G_0 \sum_{q= -\infty}^{\infty} c_{gq}(t,k_q)  (-1)^q e^{-i(\omega_q-\Delta_0)t},    \notag\\
\dot{c}_{gq}(t,k_q) = i\gamma c_{eq}(t,k_q)   \label{discretemodel4} \\
 ~~~~~~ + i\sqrt{2} \sqrt{\frac{\pi}{2L}} G_0  \sum_{q= -\infty}^{\infty}  c_g(t) (-1)^q e^{i(\omega_q-\Delta_0)t}   \notag\\
 ~~~~~~ + i  \sqrt{\frac{\pi}{2L}} G_0\sum_{k_2= -\infty}^{\infty}  (-1)^{k_2} e^{-i(\omega_{k_2}-\Delta_0)t} c_{gkk}(t,k_q,k_2)    \notag\\
 ~~~~~~ + i \sqrt{\frac{\pi}{2L}} G_0 \sum_{k_1= -\infty}^{\infty}  (-1)^{k_1} e^{-i(\omega_{k_1}-\Delta_0)t} c_{gpq}(t,k_1,k_q),   \notag\\
\dot{c}_{gpq}(t,k_p,k_q) = i\sqrt{\frac{\pi}{2L}}c_{gq}(t,k_p)G^*(k_q,t) \label{discretemodel5}\\
 ~~~~~~ + i\sqrt{\frac{\pi}{2L}}c_{gq}(t,k_q)G^*(k_p,t).  \notag
\end{numcases}
\end{subequations}
\end{scriptsize}

\begin{lemma}\label{decrease}
\cite{ScullyTreatment}~The amplitude $c_{gq}$ of the mode $k_q$ transmitted from the cavity to the waveguide is proportional to $\frac{1}{\sqrt{(ck_q-\Delta_0)^2 + \Gamma^2}}$, where $\Gamma = \frac{c(1-r)}{2l}$ with $r$ being the reflection coefficient of the mirror at $z=0$.
\end{lemma}

{\bf Lemma \ref{decrease}} has been proved in  Ref. \cite{ScullyTreatment}. A brief introduction of the interaction fields between the cavity and the waveguide of the discrete mode $k_q$ is given in {\bf Appendix~\ref{Sec:Maxwell}}. {\bf Lemma \ref{decrease}} reveals that the quantum field in the waveguide, i.e., $c_{gq}(t,k_q)$, is Lorentzian with a narrowband in the frequency domain, and the major frequency component of $k_q$ is around $\frac{\Delta_0}{c}$ because $\frac{1}{\sqrt{(ck_q-\Delta_0)^2 + \Gamma^2}}$ is peaked at $k_q = \frac{\Delta_0}{c}$. See also {\bf Appendix~\ref{Sec:Maxwell}} for more details.

\begin{assumption} \label{reflection}
The cavity's decay rate is small and the reflection rate $r$ of the mirror at $z=0$ is close to $1$.
\end{assumption}

With the selected discrete modes in Eq.~(\ref{con:discreteCoupling}) which can maximize the feedback coupling efficiency between the cavity and the waveguide, we concentrate on the feedback design with the cavity of high quality as stated in {\bf Assumption \ref{reflection}}, which is enough to induce efficient feedback when combined with Eq.~(\ref{con:discreteCoupling}), and this has been widely adopted in the feedback design, see, e.g.,  Refs.~\cite{ScullyTreatment,Geathreelevel}.

\begin{Theorem}\label{discrete}
The integral $\int_0^t  c_{gq}(u,k_q) du \approx f(t)\delta(ck_q-\Delta_0)$
when $1-r \ll \frac{2\pi l}{L}$ and the time domain envelope $f(t)$ satisfies $f'(t) \approx 0$. 
\end{Theorem}

\begin{Proof}
For the semi-transparent mirror with $|r| \leq 1$, and $1-r$ represents the field transmitted from the cavity to the waveguide. 
When $1-r \ll \frac{2\pi l}{L}$,
\begin{equation} \label{con:ckq}
\begin{aligned}
ck_{q+1} - ck_q = \frac{c\pi}{L} \gg \frac{c(1-r)}{2l} =\Gamma.
\end{aligned}
\end{equation}
Similarly, $|ck_{q-1} - ck_q|\gg \Gamma$. Hence, for the discrete modes $k_q \neq \frac{\Delta_0}{c}$, $(ck_q-\Delta_0)^2 \gg \Gamma^2$ and $\frac{1}{\sqrt{(ck_q-\Delta_0)^2 + \Gamma^2}} \approx 0$, thus the photon transmitted from the cavity to the waveguide satisfies that $c_{gq}(t,k_q) \propto \delta(ck_q - \Delta_0)$.
Denote $\int_0^t  c_{gq}(u,k_q) du = f(t)\delta(ck_q-\Delta_0)$ where the envelope $f(t)$ is the function of time. Thus
\begin{equation} \label{con:ft}
\begin{aligned}
f(t) =\frac{\int_0^t  c_{gq}(u,k_q) du}{\delta(ck_q-\Delta_0)},
\end{aligned}
\end{equation}
and its derivative $f'(t) = \frac{c_{gq}(t,k_q)}{\delta(ck_q-\Delta_0)}$. When $ck_q = \Delta_0$, $f'(t) \approx 0$ because $\delta(ck_q-\Delta_0) \gg 1$ and $c_{gq}(t,k_q)$ is finite. When $ck_q \neq\Delta_0$, by   {\bf Lemma~\ref{decrease}} we also have $f'(t) \approx 0$. \qed
\end{Proof}

{\bf Theorem~\ref{discrete}} reflects the physical fact that the coupling between the cavity and waveguide through the semi-transparent mirror located at $z = 0$ is maximized when the discrete mode in the waveguide is resonant with the cavity mode. As a result, the linewidth of the radiation field is usually narrow and rapidly decreases with the increasing of the detunning between the discrete waveguide mode and the cavity mode. More discussions can be found in Ref.~\cite{CavityEquation}.

\begin{lemma}\label{cgdiscreteproperty}
For the discrete mode control equation   (\ref{discretemodel3}),
\begin{equation} \label{con:sumEq0}
\begin{aligned}
\sum_{q= -\infty}^{\infty} c_{gq}(t,k_q)  (-1)^q e^{-i(\omega_q-\Delta_0)t} = 0,
\end{aligned}
\end{equation}
when $1-r \ll \frac{2\pi l}{L}$.
\end{lemma}
\begin{Proof}
Notice that
\begin{small}
\begin{equation} \label{con:adddot}
\begin{aligned}
&~~~\frac{d}{dt}\sum_{q= -\infty}^{\infty} \int_0^t  c_{gq}(u,k_q) du  (-1)^q e^{-i(\omega_q-\Delta_0)t}\\
&=\sum_{q= -\infty}^{\infty}  c_{gq}(t,k_q) (-1)^q e^{-i(\omega_q-\Delta_0)t} \\
&~~~~ - i(\omega_q-\Delta_0) \sum_{q= -\infty}^{\infty} \int_0^t  c_{gq}(u,k_q) du  (-1)^q e^{-i(\omega_q-\Delta_0)t}.
\end{aligned}
\end{equation}
\end{small}
We have
\begin{small}
\begin{equation} \label{con:calculation1}
\begin{aligned}
&~~~~\sum_{q= -\infty}^{\infty} c_{gq}(t,k_q)  (-1)^q e^{-i(\omega_q-\Delta_0)t} \\
&=  \left [\sum_{q= -\infty}^{\infty} \int_0^t  c_{gq}(u,k_q) du  (-1)^q e^{-i(\omega_q-\Delta_0)t}\right ]' \\
&~~~~+ i(\omega_q-\Delta_0) \sum_{q= -\infty}^{\infty} \int_0^t  c_{gq}(u,k_q) du  (-1)^q e^{-i(\omega_q-\Delta_0)t} \\
&=  \left [\sum_{q= -\infty}^{\infty} f(t) \delta(q-[\frac{\Delta_0L}{c\pi}])  (-1)^q e^{-i(\omega_q-\Delta_0)t}\right ]' \\
&~~~~+ i(\omega_q-\Delta_0) \sum_{q= -\infty}^{\infty} \int_0^t  c_{gq}(u,k_q) du  (-1)^q e^{-i(\omega_q-\Delta_0)t} \\
&= -i\sum_{q= -\infty}^{\infty}(\omega_q-\Delta_0) f(t) \delta(q-[\frac{\Delta_0L}{c\pi}])  (-1)^q e^{-i(\omega_q-\Delta_0)t}\\
&~~~~ + f'(t) \left [\sum_{q= -\infty}^{\infty} \delta(q-[\frac{\Delta_0L}{c\pi}])  (-1)^q e^{-i(\omega_q-\Delta_0)t}\right ]\\
&~~~~ +  i\sum_{q= -\infty}^{\infty}(\omega_q-\Delta_0)  f(t) \delta(q-[\frac{\Delta_0L}{c\pi}])  (-1)^q e^{-i(\omega_q-\Delta_0)t},\\
\end{aligned}
\end{equation}
\end{small}
where $[\frac{\Delta_0L}{c\pi}]$ represents the  integer  nearest to $\frac{\Delta_0L}{c\pi}$, the first and third terms cancel each other, and the second term is approximately zero when  $1-r \ll \frac{2\pi l}{L}$ as $f'(t) \approx 0$ by {\bf Theorem~\ref{discrete}}. Thus Eq.~(\ref{con:sumEq0}) holds. \qed
\end{Proof}

By {\bf Theorem~\ref{discrete}}, Eqs. (\ref{discretemodel1}) and (\ref{discretemodel2}), we get
\begin{small}
\begin{equation} \label{con:Cediscrete}
\begin{aligned}
\dot{c}_{e}(t) &=i\sqrt{2}\gamma c_g(t) - \frac{\pi}{2L} G_0^2  \sum_{q= 0}^{\infty}  c_e(t-q\tau) \tau e^{i(\Delta_0 - \frac{\pi}{\tau} )q\tau} \\
  &- \sqrt{\frac{\pi}{2L}} G_0\gamma \delta(q-[\frac{\Delta_0L}{c\pi}])  (-1)^q e^{-i(\omega_q-\Delta_0)t}.
\end{aligned}
\end{equation}
\end{small}
Please refer to {\bf Appendix~\ref{Sec:discreteproof}} for a detailed derivation of Eq. \eqref{con:Cediscrete}.

Based on Eq. \eqref{con:Cediscrete} and {\bf Lemma \ref{cgdiscreteproperty}}, we can obtain the following result.

\begin{Theorem}\label{oscillate}
In the discrete modes case, the amplitudes $c_e(t)$ and $c_g(t)$ oscillate with damping close to zero, and there are almost no two-photon states in the waveguide.
\end{Theorem}

\begin{Proof}


Denote $q^* = [\frac{\Delta_0L}{c\pi}]$. Integrating both sides of Eq. \eqref{con:Cediscrete} gives
\begin{small}
\begin{equation} \label{con:Cetdiscrete}
\begin{aligned}
c_e(t)&= i\sqrt{2}\gamma \int_0^t c_g(t)dt\\
 &- \frac{\pi}{2L} G_0^2  \sum_{q= 0}^{\infty} \int_0^t c_e(u-q\tau) \tau e^{i(\Delta_0 - \frac{\pi}{\tau} )q\tau} du  \\
 &-\sqrt{\frac{\pi}{2L}} G_0\gamma t (-1)^{q^*}. \\
\end{aligned}
\end{equation}
\end{small}
Hence, by differentiating Eq.~\eqref{con:Cetdiscrete} twice we get
\begin{small}
\begin{equation} \label{con:ddotcediscrete}
\begin{aligned}
&\ddot{c}_e(t) = -2\gamma^2c_e(t) - \frac{\pi}{2L} G_0^2  \sum_{q= 0}^{\infty}  \dot{c}_e(t-q\tau) \tau e^{i(\Delta_0 - \frac{\pi}{\tau} )q\tau}.
\end{aligned}
\end{equation}
\end{small}
Applying the Laplace transform to Eq. \eqref{con:ddotcediscrete} we get
\begin{small}
\begin{equation} \label{con:ddotcediscreteLaplace}
\begin{aligned}
&s^2C_e(s) -sc_e(0) - \dot{c}_e(0)\\
& = -2\gamma^2C_e(s) - \frac{\pi}{2L} G_0^2  \sum_{q= 0}^{\infty}  [sC_e(s)-c_e(0)]e^{-q\tau s} \tau e^{i(\Delta_0 - \frac{\pi}{\tau} )q\tau} \\
&= -2\gamma^2C_e(s) - \frac{\pi}{2L} G_0^2 \tau [sC_e(s)-c_e(0)] \sum_{q= 0}^{\infty}   e^{i[(\Delta_0 + is)\tau - \pi]q} \\
&= -2\gamma^2C_e(s) - \frac{\pi}{2L} G_0^2 \tau [sC_e(s)-c_e(0)] \\
&~~~~~\sum_{q= 0}^{\infty}  \delta[\frac{(\Delta_0+is)\tau-\pi}{2\pi} - q]\\
&= -2\gamma^2C_e(s).
\end{aligned}
\end{equation}
\end{small}
According to {\bf Assumption 1} and Eq.~(\ref{discretemodel1}), $c_e(0) =1$. Moreover, $\dot{c}_e(0) = 0$ because $c_g(0) = c_{eq}(0,k_q) = 0$. Thus, $C_e(s) \approx \frac{s}{s^2 + 2\gamma^2}$ in Eq.~(\ref{con:ddotcediscreteLaplace}). That is, $c_e(t)$  oscillates with damping being close to zero. Finally,  according to Eq. (\ref{discretemodel3}) and {\bf Lemma~\ref{cgdiscreteproperty}},  $\dot{c}_g(t) = i\sqrt{2}\gamma c_e(t)$, therefore $c_g(t)$ also oscillates with damping being close to zero. \qed
\end{Proof}

\begin{figure}[h]
\centerline{\includegraphics[width=1.1\columnwidth]{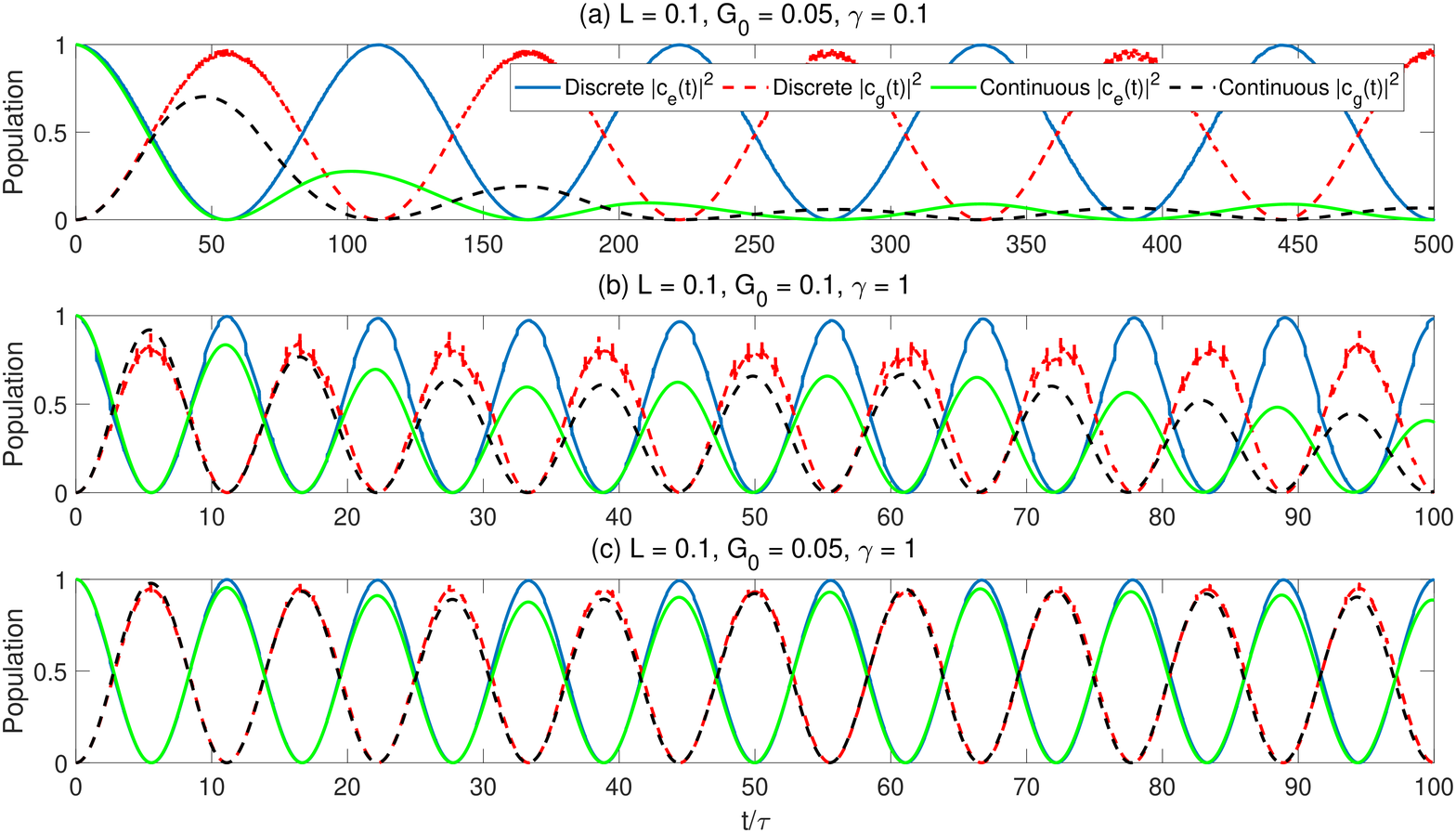}}
\caption{The comparison of the coherent feedback control with discrete and continuous coupling modes between the cavity and the waveguide.}
	\label{fig:compareCD}
\end{figure}

In the numerical simulations for the discrete coupling scheme based on Eq.~(\ref{con:PopuquationResultdiscrete}), as shown in Fig.~\ref{fig:compareCD}, $k$ is uniformly sampled as $k_q = \frac{(2q+1)\pi}{2L}$, where $q = 1,2,\dots,39$, and $L = 0.1$m.
 The simulations in Figs. \ref{fig:compareCD}(a) and (b) show that the discrete-coupling scheme can maintain the Rabi oscillation in the cavity no matter whether the coupling between the atom and cavity is weak or strong, which  agrees with {\bf Theorem~\ref{oscillate}}.  This is different from the continuous coupling scheme as shown in Eq. (\ref{con:cet7}). Moreover, the damped amplitudes of $c_g(t)$ and $c_e(t)$ in the continuous coupling scheme also show that there are generated two-photon states in the waveguide, while the oscillating $c_e(t)$ and $c_g(t)$ in the discrete coupling scheme reveal that the two photons cannot simultaneously exist in the waveguide. 
 Fig.\ref{fig:compareCD}(c) further shows that when the coupling between the atom and cavity is much larger than that between the cavity and waveguide, the atom will oscillate between its ground and excited states no matter whether the coupling between the waveguide and cavity is continuous or discrete. All the three simulations under different parameter settings have illustrated the fact that the discrete coupling scheme can maintain the Rabi oscillations of the two-level atom in the cavity.

\section{Conclusion} \label{Sec:conclusion}
In this paper, we have studied a coherent feedback control scheme in the architecture that a waveguide is coupled with a cavity containing a two-level atom. The control performance depends critically on whether the coupling mode between the cavity and the waveguide is continuous or periodically discrete. For the continuous modes case, the generation of two-photon states can be controlled by tuning the length of the waveguide as well as the coupling between the waveguide and the cavity. The populations of the two-photon states in the waveguide can be maximized when the length of the waveguide is well designed. Moreover, for the waveguide that is long enough and the round-trip delay of the feedback loop is longer than the evolution time of the quantum state, the generation rate of the two photons can be optimized by engineering the coupling strength between the waveguide and the cavity as well as that between the cavity and the atom. 

The major difference between the discrete mode scheme and the continuous mode scheme is that, the discrete mode can stablize the Rabi oscillation in the cavity and there are no stable two-photon states in the waveguide.
The results in this paper are useful for the coherent feedback design of quantum optical systems to generate multi-photon states or Rabi oscillations, and can be further applied in quantum networks for quantum information processing.

\begin{appendices}

\section{Quantum feedback control with the waveguide continuously coupled to the cavity} \label{Sec:Model}
In this appendix, we give the derivation of  Eqs. \eqref{con:cet7}-\eqref{con:cgPopu2} in the main text. The populations of the quantum state are governed by Eq.  (\ref{con:Popuquation}). As argued in the main text, using Eqs. \eqref{con:cet7}-\eqref{con:cgPopu2}, we can re-write Eq.  (\ref{con:Popuquation}) as Eq. \eqref{con:PopuquationResult}, which demonstrates clearly the influence of  the round trip delay $\tau$ on population distributions.

The rough idea of derivation is as follows. We first derive Eq.~(\ref{con:cet7}) by taking the integration of Eq.~(\ref{model2}) into Eq.~(\ref{model1}) in the main text. Then Eq.~(\ref{con:cgktnewNotau4}) can be derived by taking the integration of Eq.~(\ref{model5}) into Eq.~(\ref{model4}). Based on this, finally Eq.~(\ref{con:cgPopu2}) can be similarly derived. In the procedure, the properties of Dirac delta functions and Assumptions 2 and 3 are used.

(i) Derivation of Eq.~(\ref{con:cet7}).

Integrating both sides of Eq. \eqref{model2} and substituting the resultant  $c_{ek}$ into Eq. \eqref{model1} yields
\begin{small}
\begin{equation} \label{con:cet1}
\begin{aligned}
\dot{c}_e(t) &=i\sqrt{2}\gamma c_g(t) - \int_0^\infty \int_0^t [c_e(t') G^*(k,t') \\
&~~~~ + \gamma  c_{gk}(t',k)]G(k,t) dt'\mathrm{d}k \\
&=i\sqrt{2}\gamma c_g(t) - \int_0^\infty \int_0^t c_e(t') G^*(k,t')G(k,t) dt'\mathrm{d}k \\
&~~~~ - \int_0^\infty \int_0^t \gamma  c_{gk}(t',k)G(k,t) dt'\mathrm{d}k, \\
\end{aligned}
\end{equation}
\end{small}
where the second term on the right-hand side can be further simplified, specifically,
\begin{small}
\begin{equation} \label{con:cet2}
\begin{aligned}
&~~~ \int_0^\infty \int_0^t c_e(t') G^*(k,t')G(k,t) dt'\mathrm{d}k \\
&=G_0^2\int_0^\infty \int_0^t  \sin^2(kL)e^{-i(\omega-\Delta_0)t}e^{i(\omega-\Delta_0)t'} c_e(t')dt'\mathrm{d}k \\
&=\frac{G_0^2}{4c}\int_0^\infty \int_0^t (2-e^{i\omega\tau}-e^{-i\omega\tau})e^{-i(\omega-\Delta_0)t}e^{i(\omega-\Delta_0)t'} c_e(t')dt'\mathrm{d}\omega \\
&=\frac{G_0^2}{4c}\int_0^\infty \int_0^t (2e^{-i(\omega-\Delta_0)(t-t')}-e^{i\Delta_0\tau}e^{-i(\omega-\Delta_0)(t-t'-\tau)}\\
&~~~~-e^{-i\Delta_0\tau}e^{-i(\omega-\Delta_0)(t-t'+\tau)}) c_e(t')dt'\mathrm{d}\omega \\
&=\frac{G_0^2\pi}{2c}\int_0^t (2\delta(t-t')-e^{i\Delta_0\tau}\delta(t-t'-\tau)\\
&~~~~-e^{-i\Delta_0\tau}\delta(t-t'+\tau)) c_e(t')dt' \\
&=\frac{G_0^2\pi}{2c} (c_e(t)-e^{i\Delta_0\tau}c_e(t-\tau)\Theta(t-\tau)), \\
\end{aligned}
\end{equation}
\end{small}
where $\tau = \frac{2L}{c}$ is the round-trip delay.

On the other hand, the third term on the right-hand side of Eq. \eqref{con:cet1} reads
\begin{small}
\begin{equation} \label{con:thirdpart}
\begin{aligned}
&~~~ \int_0^\infty \int_0^t \gamma  c_{gk}(t',k)G(k,t) dt'\mathrm{d}k\\
&=\gamma \int_0^\infty \int_0^t   c_{gk}(t',k)G_0\sin(kL)e^{-i(\omega-\Delta_0)t} dt'\mathrm{d}k\\
&=\gamma G_0 e^{i\Delta_0t}\int_0^\infty \int_0^t   c_{gk}(t',k)\sin(\frac{\omega \tau}{2})e^{-i\omega t} dt'\mathrm{d}k \\
&=\frac{\gamma G_0 e^{i\Delta_0t}}{2ic}  \int_0^t \int_0^\infty  c_{gk}(t',\frac{\omega}{c}) (e^{i\omega(\frac{ \tau}{2}-t)}-e^{-i\omega(\frac{\tau}{2}+t)}) \mathrm{d}\omega dt'.
\end{aligned}
\end{equation}
\end{small}

By means of the following lemma, it can be shown that this term is 0.

\begin{lemma} \label{Integr0}
By {\bf Assumption~\ref{bound}}, we have
\[\int_0^t \int_0^\infty c_{gk}(t',\frac{\omega}{c})  (e^{i\omega(\frac{ \tau}{2}-t)}-e^{-i\omega(\frac{\tau}{2}+t)}) \mathrm{d}\omega dt' = 0.
\]
\end{lemma}

\begin{Proof}
Because $\delta(t+\tau) = 0$ for $t \geq 0$ and $\lim_{k\rightarrow \infty} c_{gk}(t,k) = 0$ as given in   {\bf Assumption~\ref{bound}}, we have
\begin{small}
\begin{equation} \label{con:cet4}
\begin{aligned}
&~~~ \int_0^t \int_0^\infty c_{gk}(t',\frac{\omega}{c})  (e^{i\omega(\frac{ \tau}{2}-t)}-e^{-i\omega(\frac{\tau}{2}+t)}) \mathrm{d}\omega dt' \\
&=\int_0^t (\delta(\frac{\tau}{2}-t)-\delta(\frac{\tau}{2}+t)) c_{gk}(t',\frac{\omega}{c})|_0^{\infty} dt'\\
 &~~~~- \int_0^t \int_0^{\infty} (\delta(\frac{\tau}{2}-t)-\delta(\frac{\tau}{2}+t)) \frac{\partial c_{gk}(t',\frac{\omega}{c})}{\partial \omega}\mathrm{d}\omega dt'\\
&=-(\delta(\frac{\tau}{2}-t)-\delta(\frac{\tau}{2}+t))\int_0^t c_{gk}(t',0)dt'\\
 &~~~~- (\delta(\frac{\tau}{2}-t)-\delta(\frac{\tau}{2}+t)) \int_0^t \int_0^{\infty}  \frac{\partial c_{gk}(t',\frac{\omega}{c})}{\partial \omega}\mathrm{d}\omega dt'\\
&=-\delta(\frac{\tau}{2}-t)\int_0^t c_{gk}(t',0)dt'
 + \delta(\frac{\tau}{2}-t) \int_0^t c_{gk}(t',0) dt'.
\end{aligned}
\end{equation}
\end{small}


Notice that $\delta(\frac{\tau}{2}-t) = 0$  for all $t \neq \frac{\tau}{2}$, and hence from Eq.~\eqref{con:cet4} we have that, for $t \neq \frac{\tau}{2}$,
\begin{equation} \label{con:cet5}
\begin{aligned}
\int_0^t \int_0^\infty c_{gk}(t',\frac{\omega}{c})  (e^{i\omega(\frac{ \tau}{2}-t)}-e^{-i\omega(\frac{\tau}{2}+t)}) \mathrm{d}\omega dt' = 0. \\
\end{aligned}
\end{equation}
Furthermore, the integrals in Eq.~(\ref{con:cet4}) are continuous when $t$ varies around $\frac{\tau}{2}$ because the evolution of amplitudes is continuous according to  {\bf Assumption~\ref{continuous}},
thus Eq.~(\ref{con:cet5}) holds when $t\geq 0$. \qed
\end{Proof}

Consequently, Eq. \eqref{con:cet1} can be written as  Eq. (\ref{con:cet7}) after combined with Eq. \eqref{con:cet2} and Eq. \eqref{con:cet5}, thus we finish the derivation of Eq. (\ref{con:cet7}).


(ii) Derivation of Eq. \eqref{con:cgktnewNotau4}.

According to Eq.~(\ref{model4}) in the main text, and together with Eqs. \eqref{model2} and \eqref{model5}, we have
\begin{small}
\begin{equation} \label{con:cgktnew}
\begin{aligned}
&~~~~\dot{c}_{gk}(t,k)\\
=&\; i\sqrt{2}c_g(t) G^*(k,t)-\gamma\int_0^t c_e(t) G^*(k,t) dt \\
 & -\int_0^t \gamma^2  c_{gk}(t,k)dt \\
 &-2\int_0^t\int_{0}^{\infty} c_{gk}(t',k)G(k',t)G^*(k',t')dk'dt' \\
 & -  2\int_0^t\int_{0}^{\infty} c_{gk}(t',k')G(k',t) G^*(k,t') \mathrm{d}k'dt'.\\
\end{aligned}
\end{equation}
\end{small}
It can be shown that
\begin{small}
\begin{equation} \label{con:integraNew}
\begin{aligned}
&~~~ \int_0^t\int_{0}^{\infty} c_{gk}(t',k)G(k',t)G^*(k',t')dk'dt' \\
&=  \int_0^t\int_{0}^{\infty} c_{gk}(t',k)G_0\sin(k'L)e^{-i(\omega'-\Delta_0)t} G_0\\
&~~~~~\sin(k'L)e^{i(\omega'-\Delta_0)t'} \mathrm{d}k'dt' \\
&= \frac{G_0^2}{c}\int_0^t\int_{0}^{\infty}c_{gk}(t',k)\sin^2(k'L)e^{-i(\omega'-\Delta_0)(t-t')}\mathrm{d}\omega'dt' \\
&=\frac{G_0^2\pi}{2c}[c_gk(t,k)-c_{gk}(t-\tau,k)e^{i\Delta_0\tau}\Theta(t-\tau)],
\end{aligned}
\end{equation}
\end{small}
and
\begin{small}
\begin{equation} \label{con:integral}
\begin{aligned}
&~~~ \int_0^t\int_{0}^{\infty} c_{gk}(t',k')G(k',t) G^*(k,t') \mathrm{d}k'dt' \\
&=  \int_0^t\int_{0}^{\infty} c_{gk}(t',k')G_0\sin(k'L)e^{-i(\omega'-\Delta_0)t} G_0 \\
&~~~~~~\sin(kL)e^{i(\omega-\Delta_0)t'} \mathrm{d}k'dt' \\
&= \frac{G_0^2\sin(kL)e^{i\Delta_0t}}{c}\int_0^t\int_{0}^{\infty}c_{gk}(t',k')\sin(k'L)\\
&~~~~~~e^{-i\omega't}e^{i(\omega-\Delta_0)t'}\mathrm{d}\omega'dt', \\
\end{aligned}
\end{equation}
\end{small}
where $\omega'=ck'$. Moreover, notice that
\begin{small}
\begin{equation} \label{con:integra2}
\begin{aligned}
&~~~ \int_{0}^{\infty}\sin(k'L)e^{-i\omega't} d\omega' \\
&=  \frac{1}{2i}\int_{0}^{\infty}(e^{i\frac{\omega'L}{c}} - e^{-i\frac{\omega'L}{c}}) e^{-i\omega't} d\omega' \\
&=  \frac{1}{2i}\int_{0}^{\infty}(e^{i\frac{\omega'\tau}{2}} - e^{-i\frac{\omega'\tau}{2}}) e^{-i\omega't} d\omega' \\
&=  \frac{1}{2i}\int_{0}^{\infty}(e^{i\omega'(\frac{\tau}{2}-t)} - e^{-i\omega'(\frac{\tau}{2}+t)} d\omega' \\
&=  \frac{1}{2i}(\delta(\frac{\tau}{2}-t) + \delta(\frac{\tau}{2}+t)).
\end{aligned}
\end{equation}
\end{small}
We have
\begin{small}
\begin{equation} \label{con:integra3}
\begin{aligned}
&~~~ \int_{0}^{\infty} c_{gk}(t',k')\sin(k'L)e^{-i\omega't} d\omega' \\
&=  \int_{0}^{\infty} c_{gk}(t',\frac{\omega'}{c})\sin(k'L)e^{-i\omega't} d\omega' \\
&=  \frac{1}{2i}(\delta(\frac{\tau}{2}-t) + \delta(\frac{\tau}{2}+t))  c_{gk}(t',\frac{\omega'}{c})|_{0}^{\infty} \\
 &~~~~ - \frac{1}{2i} \int_{kc}^{\infty} (\delta(\frac{\tau}{2}-t) + \delta(\frac{\tau}{2}+t)) \frac{\partial c_{gk}(t',\frac{\omega'}{c})}{\partial \omega'} d\omega'. \\
\end{aligned}
\end{equation}
\end{small}
Because
\begin{equation} \label{con:finitereason}
\begin{aligned}
\int_0^\infty c_{gk}(t,k)c^*_{gk}(t,k) dk \leq1,\\
\end{aligned}
\end{equation}
and
\begin{equation} \label{con:finitereason2}
\begin{aligned}
\lim_{k\rightarrow \infty} c_{gk}(t,k) = 0\\
\end{aligned}
\end{equation}
according to {\bf Assumption~\ref{bound}} in the main text,  Eq. \eqref{con:integra3} becomes
\begin{small}
\begin{equation} \label{con:EqcgkAppen}
\begin{aligned}
&\int_{0}^{\infty} c_{gk}(t',k')\sin(k'L)e^{-i\omega't} d\omega'\\
=& - \frac{1}{2i} \int_{kc}^{\infty} (\delta(\frac{\tau}{2}-t) + \delta(\frac{\tau}{2}+t)) \frac{\partial c_{gk}(t',\frac{\omega'}{c})}{\partial \omega'} d\omega'.
\end{aligned}
\end{equation}
\end{small}
Substituting Eq. \eqref{con:EqcgkAppen} into Eq. \eqref{con:integral} yields
\begin{small}
\begin{equation} \label{con:integral2}
\begin{aligned}
&~~~ \int_0^t\int_{0}^{\infty} c_{gk}(t',k')G(k',t) G^*(k,t') \mathrm{d}k'dt' \\
&= \frac{G_0^2\sin(kL)e^{i\Delta_0t}}{c}\int_0^t\int_{0}^{\infty}c_{gk}(t',k')\sin(k'L)\\
&~~~~~~e^{-i\omega't}e^{i(\omega-\Delta_0)t'}\mathrm{d}\omega'dt' \\
&= \frac{G_0^2\sin(kL)e^{i\Delta_0t}}{c} \int_0^t [-\frac{1}{2i}(\delta(\frac{\tau}{2}-t) + \delta(\frac{\tau}{2}+t)) \\
&~~~~~~ c_{gk}(t',k) - \frac{1}{2i} \int_{0}^{\infty} (\delta(\frac{\tau}{2}-t) + \delta(\frac{\tau}{2}+t)) \\
&~~~~~~\frac{\partial c_{gk}(t',k')}{\partial k'} dk']e^{i(\omega-\Delta_0)t'}dt'\\
&= -\frac{G_0^2\sin(kL)e^{i\Delta_0t}}{2ci} (\delta(\frac{\tau}{2}-t) + \delta(\frac{\tau}{2}+t))\\
&~~~~~~ \int_0^t   c_{gk}(t',k)e^{i(\omega-\Delta_0)t'}dt'\\
&~~~~-\frac{G_0^2\sin(kL)e^{i\Delta_0t}}{2ci}  (\delta(\frac{\tau}{2}-t) + \delta(\frac{\tau}{2}+t)) \\
&~~~~~~ \int_0^t \int_{0}^{\infty} \frac{\partial c_{gk}(t',k')}{\partial k'} dk'e^{i(\omega-\Delta_0)t'}dt'.\\
\end{aligned}
\end{equation}
\end{small}
Substituting Eqs. \eqref{con:integraNew} and \eqref{con:integral2} into Eq. \eqref{con:cgktnew}, we have that when $t\neq \frac{\tau}{2}$,
\begin{small}
\begin{equation}
\begin{aligned}
&\dot{c}_{gk}(t,k)\\
=&\; i\sqrt{2}c_g(t) G^*(k,t)-\gamma\int_0^t c_e(t) G^*(k,t) dt \\
&-\int_0^t \gamma^2  c_{gk}(t,k)dt - \frac{G_0^2\pi}{c}c_{gk}(t,k)\\
&+\frac{G_0^2\pi}{c} c_{gk}(t-\tau,k)e^{i\Delta_0\tau}\Theta(t-\tau).
\end{aligned}
\end{equation}
\end{small}
Because $c_{gk}(t,k)$ is second-order differentiable according to equation (\ref{con:Popuquation}), $\dot{c}_{gk}(t,k)$ is continuous, the above equation holds  when $t\geq 0$ including the time point $t=\frac{\tau}{2}$. We completed the derivation of Eq. \eqref{con:cgktnewNotau4}.

(iii) Derivation of  Eq.~(\ref{con:cgPopu2}).

We substitute Eq.~(\ref{con:cgktnewNotau4}) in the main text into Eq.~(\ref{model3}) to get
\begin{small}
\begin{equation}\label{con:Addcg}
\begin{aligned}
&\dot{c}_g(t) = i\sqrt{2}\gamma c_e(t) + i\sqrt{2}\int_0^\infty c_{gk}(t,k) G(k,t) \mathrm{d}k \\
&=i\sqrt{2}\gamma c_e(t) - 2 G_0^2\int_0^\infty \int_0^t c_g(\nu) e^{i(\omega-\Delta_0)(\nu-t)}d\nu \sin^2(kL) \mathrm{d}k\\
&~~~~-i\sqrt{2}\gamma G_0^2\int_0^\infty   \int_0^t\int_0^u c_e(\nu) e^{i(\omega-\Delta_0)(\nu-t)} d\nu du  \sin^2(kL)\mathrm{d}k \\
&~~~~-i\sqrt{2}\gamma^2 G_0\int_0^\infty \int_0^t\int_0^u  c_{gk}(\nu,k)d\nu du\sin(kL)e^{-i(\omega-\Delta_0)t} \mathrm{d}k \\
&~~~~-i\sqrt{2}\frac{G_0^3\pi}{c} \int_0^t \int_0^\infty  [c_{gk}(\nu,k)-c_{gk}(\nu-\tau,k)e^{i\Delta_0\tau}]\\ &~~~~\sin(kL)e^{-i(\omega-\Delta_0)t} \mathrm{d}k d\nu.
\end{aligned}
\end{equation}
\end{small}
When $t\neq \frac{\tau}{2}$, according to Eq.~(\ref{con:EqcgkAppen}),
\begin{small}
\begin{equation}  \label{con:cgkIntk0}
\int_0^\infty c_{gk}(\nu,k) \sin(kL)e^{-i(\omega-\Delta_0)t} dk=0,
\end{equation}
\end{small}
and
\[
\int_0^\infty c_{gk}(\nu-\tau,k)e^{i\Delta_0\tau} \sin(kL)e^{-i(\omega-\Delta_0)t} dk =0.
\]
Thus, by continuity,  the last term in Eq. \eqref{con:Addcg} is 0.

Additionally, denote $\tilde{c}_{gk}(u,k) = \int_0^uc_{gk}(\nu,k)d\nu$. Then the integrate in the third line of Eq.~(\ref{con:Addcg})
\[
\int_0^\infty \int_0^t\int_0^u  c_{gk}(\nu,k)d\nu du\sin(kL)e^{-i(\omega-\Delta_0)t} \mathrm{d}k = 0,
\]
which can be proved by replacing $c_{gk}(\nu,k)$ in Eq.~(\ref{con:cgkIntk0}) with $\tilde{c}_{gk}(u,k)$.
Hence, Eq.~(\ref{con:Addcg}) can be simplified as:

\begin{small}
\begin{align*}
&\dot{c}_g(t) 
=i\sqrt{2}\gamma c_e(t) - 2 G_0^2\int_0^\infty \int_0^t c_g(\nu) e^{i(\omega-\Delta_0)(\nu-t)}d\nu \sin^2(kL) \mathrm{d}k\\
&~~~~-i\sqrt{2}\gamma G_0^2\int_0^\infty   \int_0^t\int_0^u c_e(\nu) e^{i(\omega-\Delta_0)(\nu-t)} d\nu du  \sin^2(kL)\mathrm{d}k \\
&=i\sqrt{2}\gamma c_e(t) - \frac{G_0^2\pi}{c}(c_g(t)-e^{i\Delta_0\tau}c_g(t-\tau))\\
&~~~~-i\frac{\sqrt{2}\gamma G_0^2}{4}   \int_0^t\int_0^u  c_e(\nu) e^{i\Delta_0(t-\nu)}\int_0^\infty  e^{-i\omega(t-\nu)}\\
&~~~~(2-e^{i\omega\tau}-e^{-i\omega\tau})\mathrm{d}k d\nu du   \\
&=i\sqrt{2}\gamma c_e(t) - \frac{G_0^2\pi}{c}(c_g(t)-e^{i\Delta_0\tau}c_g(t-\tau))\\
&~~~~-i\frac{\sqrt{2}\gamma G_0^2}{4}   \int_0^t\int_0^u  c_e(\nu) e^{i\Delta_0(t-\nu)} \\
&~~~~\int_0^\infty  (2e^{-i\omega(t-\nu)}-e^{-i\omega(t-\nu-\tau)}-e^{-i\omega(t-\nu+\tau)})\mathrm{d}k d\nu du   \\
&=i\sqrt{2}\gamma c_e(t) - \frac{G_0^2\pi}{c}(c_g(t)-e^{i\Delta_0\tau}c_g(t-\tau))\\
&~~~~-i\frac{\sqrt{2}\gamma G_0^2\pi}{2c}   \int_0^t\int_0^u  c_e(\nu) e^{i\Delta_0(t-\nu)} (2\delta(t-\nu)\\
&~~~~ -\delta(t-\nu-\tau)-\delta(t-\nu+\tau)) d\nu du   \\
&=i\sqrt{2}\gamma c_e(t) - \frac{G_0^2\pi}{c}(c_g(t)-e^{i\Delta_0\tau}c_g(t-\tau))\\
&~~~~-i\frac{\sqrt{2}\gamma G_0^2\pi}{2c}   \int_{t-\tau}^t c_e(t-\tau)e^{i\Delta_0\tau} du   \\
&=i\sqrt{2}\gamma c_e(t) - \frac{G_0^2\pi}{c}(c_g(t)-e^{i\Delta_0\tau}c_g(t-\tau))\\
&~~~~-i\frac{\sqrt{2}\gamma G_0^2\pi}{2c} \tau c_e(t-\tau)e^{i\Delta_0\tau}.   \\
\end{align*}
\end{small}

Consequently,
\begin{small}
\begin{equation} \label{con:cgPopu2_new}
\begin{aligned}
&\dot{c}_g(t)\\
= &\; i\sqrt{2}\gamma c_e(t) - \frac{G_0^2\pi}{c}(c_g(t)-e^{i\Delta_0\tau}c_g(t-\tau)\Theta(t-\tau))\\
& -i\frac{\sqrt{2}\gamma G_0^2\pi}{2c} \tau c_e(t-\tau)e^{i\Delta_0\tau}\Theta(t-\tau),
\end{aligned}
\end{equation}
\end{small}
which is Eq. (\ref{con:cgPopu2}).

\section{Proof of {\bf Lemma~\ref{tinytau}}} \label{Sec:proofLemma1}
\begin{Proof}
When $\kappa \tau \ll 1$, namely $L\ll 1$, the parameters defined in Eqs. \eqref{con:jan8_1}-\eqref{con:jan8_2} satisfy that $E =R = \kappa(\cos(\Delta_0\tau)-1) + i\kappa \sin(\Delta_0\tau)\approx 0$, $D = \frac{\sqrt{2}}{2} [\Delta_0-\omega-\sin(\Delta_0\tau) + i\kappa(\cos(\Delta_0\tau) -1)]\approx \frac{\sqrt{2}}{2} [\Delta_0-\omega-\sin(\Delta_0\tau)] \approx\frac{\sqrt{2}}{2}(\Delta_0-\omega)$. Thus $\sqrt{R^2 - \gamma^2} \approx i\gamma$, $\gamma \ll \Delta_0$, $F\approx 0$ in Eq. \eqref{con:jan8_1}. Consequently,
\begin{small}
\begin{equation} \label{con:HIJK}
\begin{aligned}
H(\omega) &= \lim_{s\rightarrow -i\gamma} \frac{s-E-i(\omega-\Delta_0+F+ \frac{2\sqrt{2}}{3}D)}{(s-i\gamma)\{[s-E-i(\omega-\Delta_0+F)]^2 +2\gamma^2\}} \\
& \approx \frac{\gamma+\frac{1}{3}(\omega-\Delta_0)}{2\gamma [-(\gamma+\omega-\Delta_0)^2 +2\gamma^2] },\\
I(\omega) &= \lim_{s\rightarrow i\gamma} \frac{s-E-i(\omega-\Delta_0+F+ \frac{2\sqrt{2}}{3}D)}{(s+i\gamma)\{[s-E-i(\omega-\Delta_0+F)]^2 +2\gamma^2\}} \\
& \approx \frac{\gamma -\frac{1}{3}(\omega-\Delta_0)}{2\gamma [-(\gamma-\omega+\Delta_0)^2 +2\gamma^2] }, \\
J(\omega) &= \lim_{s\rightarrow E+i(\omega-\Delta_0+F) - i\sqrt{2}\gamma} \\
&~~~~\frac{s-E-i(\omega-\Delta_0+F+ \frac{2\sqrt{2}}{3}D)}{(s^2+\gamma^2)[s-E-i(\omega-\Delta_0+F) - i\sqrt{2}\gamma]} \\
& = \frac{\sqrt{2}\gamma +\frac{2}{3}(\Delta_0-\omega)}{2\sqrt{2}\gamma[ -(\omega-\Delta_0-\sqrt{2}\gamma)^2 +\gamma^2 ] } ,\\
K(\omega) &= \lim_{s\rightarrow E+i(\omega-\Delta_0+F) + i\sqrt{2}\gamma} \\
&~~~~\frac{s-E-i(\omega-\Delta_0+F+ \frac{2\sqrt{2}}{3}D)}{(s^2+\gamma^2)[s-E-i(\omega-\Delta_0+F) + i\sqrt{2}\gamma]} \\
& = \frac{\sqrt{2}\gamma - \frac{2}{3}(\Delta_0-\omega)}{2\sqrt{2}\gamma[ -(\omega-\Delta_0+\sqrt{2}\gamma)^2 +\gamma^2 ] }. \\
\end{aligned}
\end{equation}
\end{small}
In summary,
\begin{small}
\begin{equation} \label{con:omega12}
\left\{
\begin{aligned}
&H(\omega) =I(2\Delta_0-\omega)^*,\\
&J(\omega) =K(2\Delta_0-\omega)^*. \\
\end{aligned}
\right.
\end{equation}
\end{small} \qed
\end{Proof}

\section{Quantum feedback control for the discrete method scheme} \label{Sec:discreteproof}
Eq (\ref{con:Cediscrete}) is proved as follows. Integrating both sides of Eq. (\ref{discretemodel2}) in the main text yields
\begin{small}
\begin{equation} \label{con:ceq}
\begin{aligned}
c_{eq}(t,k_q) = &i\sqrt{\frac{\pi}{2L}} G_0 (-1)^q \int_0^t c_e(u) e^{i(\omega_q-\Delta_0)u} du \\
&+ i\gamma \int_0^t  c_{gq}(u,k_q) du.
\end{aligned}
\end{equation}
\end{small}
Substituting Eq. \eqref{con:ceq} into Eq. \eqref{discretemodel1} gives
\begin{small}
\begin{equation} \label{con:cetdot}
\begin{aligned}
&\dot{c}_{e}(t) \\
=&\; i\sqrt{2}\gamma c_g(t) \\
&- \frac{\pi}{2L} G_0^2 \sum_{q= -\infty}^{\infty} \int_0^t c_e(u) e^{i(\omega_q-\Delta_0)(u-t)} du \\
&- \sqrt{\frac{\pi}{2L}} G_0\gamma \sum_{q= -\infty}^{\infty} \int_0^t  c_{gq}(u,k_q) du  (-1)^q e^{-i(\omega_q-\Delta_0)t}.
\end{aligned}
\end{equation}
\end{small}
Notice that
\begin{small}
\begin{equation} \label{con:cetComponent2}
\begin{aligned}
&~~~~ \sum_{q= -\infty}^{\infty} e^{i(\omega_q-\Delta_0)(u-t)} \\
& =  e^{i(\frac{\pi}{\tau} -\Delta_0)(u-t)}   \sum_{q= -\infty}^{\infty} e^{i(\omega_q-\frac{\pi}{\tau})(u-t)}\\
& =  e^{i(\frac{\pi}{\tau} -\Delta_0)(u-t)}   \sum_{q= -\infty}^{\infty} e^{i(\frac{(2q+1)\pi}{\tau} -\frac{\pi}{\tau})(u-t)}\\
& =  e^{i(\frac{\pi}{\tau} -\Delta_0)(u-t)}   \sum_{q= -\infty}^{\infty} e^{i2q\pi\frac{u-t}{\tau}}.\\
\end{aligned}
\end{equation}
\end{small}
According to the property of the Dirac comb,
\[
\frac{1}{\tau}  \sum_{q= -\infty}^{\infty} e^{i2\pi q\frac{u-t}{\tau}} = \sum_{q= -\infty}^{\infty} \delta(u-t-q\tau),
\]
Eq. \eqref{con:cetComponent2} becomes
\begin{small}
\begin{equation} \label{con:cetComponent3}
\begin{aligned}
&\sum_{q= -\infty}^{\infty} e^{i(\omega_q-\Delta_0)(u-t)} \\
=&\;  \tau e^{i(\frac{\pi}{\tau} -\Delta_0)(u-t)}  \sum_{q= -\infty}^{\infty} \delta(u-t-q\tau).\\
\end{aligned}
\end{equation}
\end{small}
As a result,
\begin{small}
\begin{equation} \label{con:cetComponentSolution}
\begin{aligned}
&~~~~\sum_{q= -\infty}^{\infty} \int_0^t c_e(u) e^{i(\omega_q-\Delta_0)(u-t)} du \\
&=\int_0^t c_e(u) \sum_{q= -\infty}^{\infty} e^{i(\omega_q-\Delta_0)(u-t)} du\\
&=\int_0^t c_e(u) \tau e^{i(\frac{\pi}{\tau} -\Delta_0)(u-t)}  \sum_{q= -\infty}^{\infty} \delta(u-t-q\tau) du\\
&= \sum_{q= -\infty}^{\infty} \int_0^t c_e(u) \tau e^{i(\frac{\pi}{\tau} -\Delta_0)(u-t)}  \delta(u-t-q\tau) du \\
&= \sum_{q= -\infty}^{0}  c_e(t+q\tau) \tau e^{i(\frac{\pi}{\tau} -\Delta_0)(t+q\tau-t)}   \\
&= \sum_{q= 0}^{\infty}  c_e(t-q\tau) \tau e^{i(\Delta_0 - \frac{\pi}{\tau} )q\tau}.   \\
\end{aligned}
\end{equation}
\end{small}

On the other hand, according to {\bf Theorem~\ref{discrete}},
\begin{small}
\begin{equation} \label{con:cetComponent3lim}
\begin{aligned}
&~~~~\sum_{q= -\infty}^{\infty} \int_0^t  c_{gq}(u,k_q) du  (-1)^q e^{-i(\omega_q-\Delta_0)t} \\
& = \delta(q-[\frac{\Delta_0L}{c\pi}])  (-1)^q e^{-i(\omega_q-\Delta_0)t}.
\end{aligned}
\end{equation}
\end{small}
By means of Eqs. \eqref{con:cetComponentSolution}-\eqref{con:cetComponent3lim},  $\dot{c}_{e}(t)$ in Eq. \eqref{con:cetdot}  can be re-written as Eq. (\ref{con:Cediscrete}) in the main text.

\section{The coupling scheme between the cavity and waveguide} \label{Sec:Maxwell}

The electromagnetic field with discrete modes in the waveguide and cavity can be represented as~\cite{ScullyTreatment}:
\begin{equation} \label{eq:electricfield}
E(z,t) = \sum_q (\zeta_q(t) U_q(z) + \zeta_q^*(t) U_q^*(z) ),
\end{equation}
where $\zeta_q(t)$ is the amplitude of the field of the mode $k_q$. Because the electromagnetic field in the waveguide and cavity are standing waves, the mode function $U_q(z)$ with $\xi_q^2 = 1$ can be represented as~\cite{CavityEquation,ScullyTreatment,spacetime,Geathreelevel}:
\begin{small}
\begin{equation} \label{con:window}
    U_q(z)=
   \begin{cases}
   \xi_q\sin k_q(z+L)&z<0,\\
   M_q\sin k(z-l)&z>0,
   \end{cases}
  \end{equation}
\end{small}
where $\xi_q$ and $M_q$ represent the amplitudes of the coupled mode $k_q$ in the  waveguide and cavity, which are divided by the semitransparent mirror at $z = 0$.
$U_q(z)$ is governed by the Maxwell equation
\begin{equation} \label{eq:Maxwell}
\frac{d^2 U_q(z)}{dz^2} + [1+\eta\delta(z)]k_q^2 U_k(z) = 0
  \end{equation}
with $\eta$ being the transmissivity of the mirror at $z=0$, and the boundary conditions are
\begin{small}
\begin{equation} \label{con:boundarycondition}
\left\{
\begin{aligned}
&U_q(0^+) = U_q(0^-),\\
&U_q(z)|_{z=-L,l} = 0, \\
&U_q'(0^+) - U_q'(0^-) = -\eta\Delta_0^2U_q(0),
\end{aligned}
\right.
\end{equation}
\end{small}
where the resonant frequency of the cavity $\omega_c = \Delta_0 \gg 1$. Solving the Maxwell equation \eqref{eq:Maxwell} with $U_q(z)$ in Eq. \eqref{con:window} gives the following equations at $z=0$:
\begin{small}
\begin{equation} \label{con:boundarycondition2}
\left\{
\begin{aligned}
&-M_q\sin(k_ql) = \xi_q\sin(k_qL), \\
&M_q\cos(k_ql) - \xi_q\cos(k_qL) = -\eta\Delta_0^2\xi_q\sin(k_qL).
\end{aligned}
\right.
\end{equation}
\end{small}
As a result,
\begin{equation} \label{con:dot0}
\begin{aligned}
&~~~~\tan(k_qL) =\frac{\tan(k_ql)}{\eta\Delta_0^2 \tan(k_ql) -1}.
\end{aligned}
\end{equation}
Using the boundary conditions in Eq. \eqref{con:boundarycondition}, the feedback coupling strength can be evaluated with the amplitude of the field in the cavity which is induced by the unit field in the waveguide as
\begin{equation} \label{con:tansin}
\begin{aligned}
\frac{M_q^2}{\xi_q^2} = \frac{\sin^2(k_qL)}{\sin^2(k_ql)} = \sin^2(k_qL) \frac{\tan^2(k_ql) +1}{\tan^2(k_ql)}
\end{aligned}
\end{equation}
and $\xi_q^2 = 1$.

Obviously, the feedback coupling is maximized when $\sin^2(k_qL) =1$. Thus in the coherent feedback control with discrete coupled modes discussed in {\bf Sec.~\ref{Sec:discrete}}, we choose the discrete modes as $k_q = \frac{(2q+1)\pi}{2L}$. See also the early illustration on the set of discrete modes of the cavity with small leakage in Ref.~\cite{fox1961resonant}.
%
%

Combining Eq.~(\ref{con:dot0}) and Eq.~(\ref{con:tansin}) with $\sin^2(k_qL) = \frac{\tan^2(k_qL)}{\tan^2(k_qL) +1}$, yields
\begin{small}
\begin{equation} \label{con:ABk}
\begin{aligned}
\frac{M_q^2}{\xi_q^2} &= \frac{\tan^2(k_ql) +1}{\tan^2(k_ql) + [\eta\Delta_0^2 \tan(k_ql)-1]^2}\\
&\approx \frac{c\Gamma/l}{(ck_q-\Delta_0)^2 +\Gamma^2},
\end{aligned}
\end{equation}
\end{small}
where $\Gamma = \frac{c(1-r)}{2l}$ with $r$ being the reflection coefficient of the mirror at $z=0$. Crudely speaking,  Eq. \eqref{con:ABk} means that the transmitted  field from the waveguide to the cavity is Lorentzian. Thus, when $ck_q-\Delta_0 = 0$, the transmitted field is maximized. More details can be found in Ref.~\cite{ScullyTreatment}. 

The electromagnetic field $E(z,t)$ can be quantized in the waveguide and cavity respectively. The field amplitude of the mode $k_q$ in the waveguide at $z < 0$ can be quantized as the operator $d_{q}$, and the field in the cavity at $z>0$ can be quantized as the operator $a$, as in Eq.~(\ref{con:discreteHamiltonian}). The coupling strength of the mode $k_q$ between the cavity and the waveguide $G_q(t)$ is equivalent with $\zeta_q(t) \sqrt{\frac{c\Gamma/l}{(ck_q-\Delta_0)^2 +\Gamma^2}}$ according to the above analysis. Thus the single photon amplitude $c_{gq}(t,k_q) \propto \frac{1}{\sqrt{(ck_q-\Delta_0)^2 + \Gamma^2}}$. Then generalized from Eq.~(\ref{con:ABk}), when the discrete mode $\omega_q = ck_q=\Delta_0$ in Eq.~(\ref{con:discreteCoupling}), the coupling between the cavity and the waveguide is maximized, as shown in {\bf Lemma \ref{decrease}} in the main text, the single photon amplitude is maximized at the Lorentzian peak around $\Delta_0$.

\end{appendices}

\end{document}